# Micropublications: a Semantic Model for Claims, Evidence, Arguments and Annotations in Biomedical Communications

Tim Clark[1,2,3]*, Paolo N. Ciccarese[1,2], Carole A. Goble[3]

[1] Department of Neurology, Massachusetts General Hospital
  55 Fruit Street, Boston MA 02114, USA

[2] Harvard Medical School
  25 Shattuck Street, Boston, MA 02115 USA

[3] School of Computer Science, University of Manchester
  Oxford Road, Manchester M13 9PL, UK

* Corresponding author

Email addresses:
    TC: tim_clark@harvard.edu; clarkt@cs.manchester.ac.uk
    PNC: paolo.ciccarese@gmail.com
    CAG: carole.goble@cs.manchester.ac.uk

# Abstract

### 1.1.1 Background

Scientific publications are documentary representations of defeasible arguments, supported by data and repeatable methods. They are the essential mediating artifacts in the ecosystem of scientific communications; the institutional "goal" of science is publishing results. The linear document publication format, dating from 1665, has survived transition to the Web.

Intractable publication volumes; the difficulty of verifying evidence; and observed problems in evidence and citation chains suggest a need for a web-friendly and machine-tractable model of scientific publications. This model should support: digital summarization, evidence examination, challenge, verification and remix, and incremental adoption. Such a model must be capable of expressing a broad spectrum of representational complexity, ranging from minimal to maximal forms.

### Results

The Micropublications semantic model of scientific argument and evidence provides these features. Micropublications support natural language statements; data; methods and materials specifications; discussion and commentary; challenge and disagreement; as well as allowing many kinds of statement formalization.

The minimal form of a micropublication is a statement with its attribution. The maximal form is a statement with its complete supporting argument, consisting of all relevant evidence, interpretations, discussion and challenges brought forward in support of or opposition to it. The spectrum of forms between these two endpoints supports nine significant use cases in nine major activity complexes we identified in the biomedical communications ecosystem.

Micropublications may be connected into deep multimodal citation / evidence networks across large corpora. They have the advantage of a transparent internal structure opening their foundational evidence to view. These networks themselves may be published as micropublications. Micropublications may be formalized and serialized in multiple ways, including in RDF. They may be added to publications as stand-off metadata.

An OWL 2 vocabulary for micropublications is available at http://purl.org/mp.

### Conclusion

We suggest that micropublications, generated by useful software tools supporting such activities as writing, editing, reviewing, and discussion, will be of great value in improving the quality and tractability of biomedical communications.

Micropublications, because they model evidence and allow qualified, nuanced assertions, can play essential roles in the scientific communications ecosystem in places where simpler, formalized and purely statement-based models, such as the nanopublications model, will not be sufficient. At the same time they will add significant value to, and are intentionally compatible with, statement-based formalizations.

# Keywords

# 1    Introduction

## 1.1    General Motivation for the Model

### 1.1.1    The Incomplete Transition to the Web and its Problems

During the past two decades the ecosystem of biomedical publications has moved from a print-based to a mainly Web-based model.  However, this transition brings with it many new problems, in the context of an exponentially increasing, intractable volume of publications  [1, 2]; of systemic problems relating to valid (or invalid) citation of scientific evidence [3, 4]; rising levels of article retractions [5, 6] and scientific misconduct [7]; of uncertain reproducibility and re-usability of results in therapeutic development [8], and lack of transparency in research publication [9].  While we now have rapid access to much of the world's biomedical literature, our methods to organize, verify, assess, combine and absorb this information in a comprehensive way, and to move discussion and annotation activities through the ecosystem efficiently, remain disappointing.

Computational methods previously proposed as solutions include ontologies [10]; textmining [11]; [12];[2]; databases [13]; knowledgebases [14]; visualization [15]; new forms of publishing [16]; digitial abstracting [1]; semantic annotating [17]; and combinations of these approaches. However, we lack a common means to orchestrate these methods. We propose to accomplish this with a layered metadata model of scientific argumentation and evidence.

### 1.1.2    Orchestration of Computation via a  Common Metadata Model

A common metadata representation of scientific claims, argument, evidence and annotation in biomedicine, should serve as an integrating point for the original publication, subsequent annotations, and all other computational methods, supporting a single framework for activities in the nine point cycle of authoring-publishing-consumption-reuse we discuss in Section 2. This cycle can be thought of as an information value chain in science.  This means that each set of disparately motivated and rewarded activities, carried out by various actors, creates and passes along value to the next, which consumes this value-added product as an input. A metadata representation to support this value chain would need to:

- serve as a common Web-friendly nucleus for value-addition and extraction across the biomedical communications ecosystem: understood, operated upon and exchanged by humans and by computers, as supplements to the linear documents they characterize;
- enable more powerful use and sharing of information in biomedicine, particularly through integration and mashup to provide the most relevant views for any social unit of researchers;
- enable the addition of value to the content while providing a detailed provenance of what was done;
- support computational processing in a way that complete papers in un-augmented linear natural language cannot yet integrate well with existing linear textual representations.

This paper introduces the Micropublications semantic metadata model. The Micropublications model is adapted to the Web, and designed for (a) representing the key arguments and evidence in scientific articles, and (b) supporting the "layering" of annotations and various useful formalizations upon the full text paper.

The present model responds to nine use cases outlined in section 1.2, in which digital summarization of scientific argumentation with its evidence and methodological support is required. These use cases, for the most part, deal directly with the scientific literature, rather than its processed reflection in curated topical databases. They illustrate how and why currently proposed "statement-based" approaches need richer representation and how this model can play such a role.

In this paper we present

- a Use Case analysis mapped to sets of common activities in the biomedical communications ecosystem, showing the potential value addition and path to implementation of the proposed model for each Use Case;
- Illustrative examples for each Use Case of how the model can be instantiated;
- a proposed Web-friendly representation using community ontologies serialized in the W3C Web Ontology Language; and
- a set of examples in RDF; with
- notes on an interface to the nanopublications model and other statement formalizations.

### 1.2  Beyond Statement-based Models

Statement-based models have been proposed as mechanisms for publishing key facts asserted in the scientific literature or in curated databases in a machine processable form. Examples include: Biological Expression Language (BEL) statements [18]; SWAN, a model for claims and hypotheses in natural language developed for the annotation of scientific hypotheses in Alzeimers Disease research [14, 19-21]; and nanopublications [22-26], which contribute to the Open PHACTS linked data warehouse of pharmacological data [26].

What we mean by "statement-based" is that they confine themselves to modeling statements found in scientific papers or databases, with limited or no presentation of the backing evidence for these statements. Some offer statement backing in the form of other statements in the scientific literature, but none actually has a complete representation of scientific argument including empirical evidence and methods. Of the three examples we mention,

- Nanopublications models only the indicated statement;

- SWAN models  a principal statement, or "hypothesis", with supporting statements, or "claims", from the same publication only, and backing references for the supporting statements;

- BEL and SWAN model backing statements from other publications in the literature by citing whole publications,   leaving the reader to determine precisely where in the cited document a backing statement actually resides;

- None of these models provide a means to build claim networks of arbitrary depth.

- None of these models provide a means to transitively close claim lineages to underlying empirical evidence – because they don't represent it.

Table 1 compares these three statement-based models.

|  | **SWAN** | **nanopublications** | **BEL** |
|---|---|---|---|
| **first specification** | 2008 [20] | 2010 [24] | 2011 [18] |
| **current specification** | 2012 [19] | 2012 [25] | 2011 |
| **statements in** | natural language | formal language | formal language |
| **statement provenance** | SWAN-PAV ontology | SWAN-PAV ontology | SWAN-PAV ontology |
| **backing references from literature** | yes | yes | yes |
| **support claim networks** | no | no | no |
| **direct empirical scientific evidence** | no | no | no |

**Table 1: A comparison of SWAN, nanopublications and Biological Expression Language.**

**Figure 1** shows a nanopublication which attempts to express the assertion from Spilman et al. 2010 [27] that "inhibition of mTOR by Rapamycin can slow or block AD progression in a transgenic mouse model of the disease". Nanopublications distill content as a graph of assertions associated with (a) provenance of the article or dataset from whence they came; and (b) a set of terms for indexing and filtering in order to identify auxiliary information in large data sets. Although this last point is represented by a named graph called "Support", this is not intended to represent argumentative support or evidence, but rather descriptive information (cell type, species, etc.) "to enable first pass filtering over large nanopublication sets" [25]. Note that formalization of the np:Assertion is somewhat awkward in this example, and requires multiple level of reification. Yet the np:Assertion is not modelling a markedly complex scientific claim.

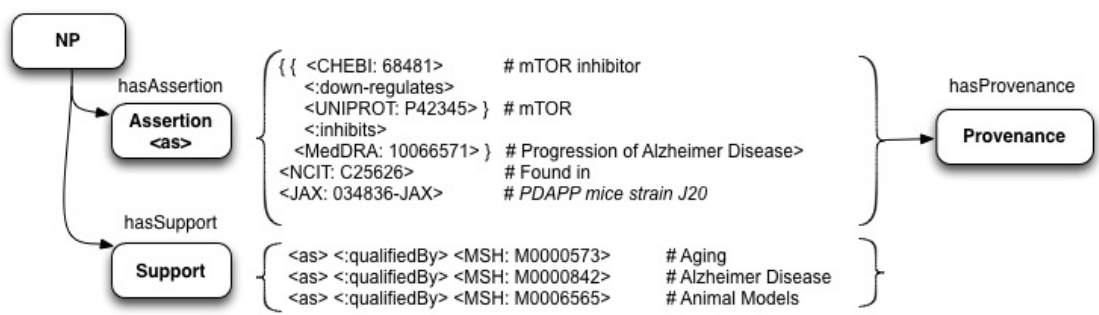

**Figure 1. Representation of statements and evidence in a nanopublication format. Nanopublications include a named graph called "Support", but this is actually a set of Qualifiers used for filtering, rather than the supporting evidence and citations. <CHEBI: 68481>, <UNIPROT: P42345>, etc., are abbrevations for the respective fully-qualified URIs.**

The intent of Statement-based models is to be relatively simple and useful for specific tasks. In the case of nanopublications, this particular model is currently presented (on a technical level) mainly for data integration across chemical and biological databases. For example, in the current nanopublication guidelines [25] a nanopublication is declared to be "a layer on top of RDF encoded data to provide a standard for the identification of individual scientific assertions within a dataset [which] enables the provenance to be assigned to each assertion and the entire dataset itself". There is no suggestion in the current specification that nanopublications may be applied directly to ordinary scientific articles, nor that they are designed to present primary scientific evidence – although more expansive claims have been made elsewhere in the literature [22, 23]. Furthermore, the fact that formalization of assertions is required, is likely an impediment to such direct use. Consequently a more comprehensive model is needed to be applied successfully across the entire ecosystem of biomedical communications.

The Micropublication approach goes beyond statements and their provenance, proposing a richer model in order to account for a more complete and broadly useful view of scientific argument and evidence, beyond that of naked assertions, or assertions supported only by literature references. It is also designed to be readily compatible with assertions coded in BEL or as nanopublications, as these models are considered useful in certain applications and will need to be integrated.

### 1.2.1    The role and importance of empirical evidence

Empirical evidence is required in scientific publications so that the scientific community may make and debate judgments based on the "interpretation of nature" rather than interpretation of texts [28]. The process of establishing new "facts" and either supplementing or overthrowing old ones, is the central work of biomedical research. Scientific assertions do not become matters of fact until the facts have been established, through judgements made over time in a complex social process. This process includes collective investigation and assessment, and may involve controversy, uncertainty, overthrow of settled opinion, and gradual convergence over time on a "current best explanation" [29]. It takes place as researchers present  arguments in the professional literature; with supporting observational data, interpretations, and theoretical and methodological context,  for evaluation by a "jury of their peers". Once a matter of fact is established, the scientific literature persists as an open documentary record, which from time to time may be challenged and reassessed.  Thus, to usefully model facts *in the process of formation*, empirical evidence, as well as formal statements and their provenance, must be a part of our model.

For example, consider the following "nano-publishable" fact given in [23, 24]: "mosquitos transmit malaria". This example reflects old science that is not currently under examination or contention, as the role of some species of *Anopheles* as malarial vectors has been well-established for over a century (roughly since the period leading up to Ronald Ross's 1902 Nobel Prize in Medicine). However, previously, in the late nineteenth century, the existence and nature of malarial vectors was an open research question [30]. Open research questions require *presentation of evidence* to establish a *warrant for belief* [31]. Thus, for  scientists working on malaria over a century ago, a statement about supposed malarial vectors without supporting empirical evidence, would not have been robust enough to enable evaluation,  and thus could not have motivated reasoned belief.

Recent results spotlight concerns with the communication of evidence and its citation.  Begley and Ellis recently found that only 11% of research findings they examined from the academic literature could be reproduced in a biopharmaceutical laboratory [8].  Fang et al. reviewed all retractions indexed in PubMed, finding that over two thirds were due to misconduct  [7]. Retractions themselves are an increasingly common event [6].  Greenberg conducted a citation network analysis of over 300 publications on a single neuromuscular disorder, and found extensive progressive distortion of citations, to the extent that reviews in reputable journals presented statements as "facts", which were ultimately based on no evidence all [3, 4].  Simkin and Roychowdhury showed that, in the sample of publications they studied, a majority of scientific citations were merely copied from the reference lists in other publications [32, 33]. The increasing interest in direct data citation of datasets, deposited in robust repositories, is another result of this growing concern with the evidence behind assertions in the literature [34].

We have incorporated a number of features in our model to enable presentation of empirical scientific evidence; therefore including data, not just assertions, as information supporting a statement; as well as other required features for scientific discourse.

### 1.2.2    The importance of natural language

As useful as formal language representations may be, any requirement that statements must only be expressed in formal language such as we find in BEL, nanopublications, and some other approaches, is a potential barrier to adoption in the publication ecosystem. We can expect to encounter scientific claims in their native environment, the biomedical literature, as relatively nuanced *arguments for qualified claims supported by evidence*. This evidence consists of citations to the literature, and novel data with supporting methods. Scientific claims "in the wild" are almost always extensively hedged or qualified, based on recognition of their incomplete or tentative nature [35]. Moreover, ordinary scientific workers present their conclusions in natural language and will continue to do so. Previous experiments such as "structured digital abstracts" have faltered: authors have little incentive to formalize their claims, only to publish them [36, 37], and tooling support is poor for those that have that desire.

Consequently the Micropublication Model must capture the natural language of claims as they appear in the literature. We treat formalization separately as an optional curatorial step. This more comprehensive approach deals with both with scientific statements "in the wild", and with the evidence that supports them, and is also compatible with statement-based formalization patterns such as nanopublications.

### 1.3    Formalizing Scientific Publications as Arguments

Our model is based on understanding scientific publications as arguments, which present a narrative of experiments or observations, the data obtained, and a reasoned interpretation ("finding") of the data's meaning [38]. Such arguments present a line of reasoning, to a "best" explanation of the data ("abductive reasoning", "inference to the best explanation", "ampliative inference") [39-42] taken in context with the published findings of others in the field.

The determination whether or not a finding is correct, is made over time by the community of the researcher's peers. What claims are considered true, may evolve over time, based on re-examination of evidence and development of new evidence. Assertions may be criticised and refuted. Thus scientific reasoning is *defeasible* [43].

Toulmin's classic model of defeasible reasoning [44], updated by Bart Verheij [45, 46], focuses on the internal structure of argument: what the author states, how it is qualified, how the author backs it up, and what other arguments may contradict it. It is a mainstream model of argument in the Artificial Intelligence (AI) community. In our model, we extend both the "support" and the "contradiction" or "rebuttal" part of this model to interargumentation, or argumentation frameworks, a topic with its own extensive literature in AI (see, e.g., [47-53], etc.). The Micropublication Model is grounded in Toulmin-Verheij; and is consistent with recent work in AI on defeasible argumentation [43, 45, 48, 50, 51, 54].

Our model provides a framework to support extensively qualified claims in natural language, as generally presented by researchers in their primary publications. Most fundamentally it adds support relations to claims in the literature, to assist in resolving primary scientific evidence within a "lineage" of assertions. Support relations, structured as graphs, back up assertions with the data, context and methodological evidence which validates them.

Micropublications permit scientific claims to be formulated minimally as any statement with an attribution (basic provenance), and maximally as entire knowledgebases with extensive evidence graphs.

Thus micropublications in their minimal form subsume or encompass statement-based models, while allowing presentation of evidential support for statements and natural language assertions as backing for formalisms.

This subsumption capability is discussed in Section 5.3.

### 1.4    Rationale for Selection  of Exemplar Publications

Our work builds particularly from experiences in applying the SWAN biomedical discourse ontology [20] to use cases in Alzheimer Disease (AD) research. SWAN led to our development of the Domeo Web annotation toolkit [17, 55], and the Annotation Ontology (AO) [56, 57] so that formal models of discourse could be readily contextualized directly in ordinary publications, as overlaid metadata.    Domeo is currently used to annotate specific reagents in biomedical publications [58], as part of the Neuroscience Information Framework (NIF), among other uses [59]. Other driving problems come from pharmaceutical research, scientific publishing, direct data archiving [60] and data citation [61].

The Micropublications Model is introduced here using some exemplar case study publications from neuroscience, specifically Alzheimer Disease (AD) research.

The AD area is perhaps emblematic of the role of competing hypotheses and alternative explanations in biomedical research. There is still not one single accepted explanatory hypothesis, and even what is probably the leading model of AD etiopathology, the "Amyloid Cascade Hypothesis" [62] has seen recent challenges, e.g. [63-66].   AD according to some authors is a syndrome rather than a single disorder [67].  There is still no cure for AD, and simple aging is still the greatest known risk factor. Given this background and our prior experience in developing an AD knowledgebase [14, 68] and associated ontology of biomedical discourse [19, 20] designed to deal with this kind of scientific conflict, we selected our exemplars from AD research.

A recent article by Spilman et al. [27] and its supporting and related materials, are rather typical of scholarly publications from this field. The principal claim in Spilman et al., is that inhibition of the mTOR pathway by rapamycin, in mice genetically engineered as models of Alzheimer Disease (AD), reverses AD pathology and symptomatology. This claim is supported in part by references to publications said to demonstrate that rapamycin inhibits mTOR [69]; and that the PDAPP mouse model used in the experiments is indeed a reasonable model of AD [70, 71]. The final pieces of support for the claim are (a) data provided by the authors, on Rapamycin-fed vs. control mice; and (b) the methods they used to obtain the data, including the feeding protocol [27],  the Morris Water Maze protocol [72, 73], and the engineered mice themselves [70, 71].

We use this and related articles as exemplars in our presentation and discussion.

### 1.5    Situating the Model in the Knowledge Value Chain

We begin our discussion by presenting the model's place in the biomedical communications knowledge or information value chain [74-76]. We do this by outlining a set of use cases in that value chain, to which the model is meant to respond.

We then present the model abstractly, followed by worked-out application of the model to each use case. The essence of this multi-use-case approach is to distribute labor costs of applying the model across the ecosystem. These costs must be aligned with direct benefits accruing to those doing the labor; while passing along "side-effect" benefits to others in the value chain who use the model.

To take one example, researchers are quite familiar with bibliographic reference management tools, and typically use them to keep track of the works they cite. These systems can include comments. An important missing piece in many of these systems is the ability to mark the specific passages cited so they can be recalled later. Core elements of the model could readily be included in such systems to keep track of specific passages and link them their supporting evidence and supporting references. This would be a value addition for the users, and elementary micropublications could be produced as side effects in the activity of bibliography management.

## 2  Use Cases

The goal of the Micropublication model is to better adapt scientific publications to production and use on the Web, in the context of the new forms either made available or required. The model supports nine main Activities in our analysis, within nine Use Case requirement sets. **Table 2** shows the mapping of Activities to Use Cases. **Figure 2** shows how each of these Use Cases, and their inputs and outputs, are situated in an Activities Lifecycle, across the biomedical communications ecosystem. This lifecycle is a cyclical value chain. We will show in this section of this article how the micropublications model can effectively support information creators and consumers (*tool users*) in this ecosystem, for important unsatisfied use cases.

| Use Case ↓ | Activity → | Author | Review | Edit & Publish | DB/KB Curate | Search for Info | Read | Discuss | Evaluate & Integrate | Design & Execute Experim't |
|---|---|---|---|---|---|---|---|---|---|---|
| **1 - Build and Use Citable Claims** | | X | X | X | X | X | X | X | X | |
| **2 - Model Evidence for Claims** | | X | X | X | X | X | X | X | X | X |
| **3 - Analyze Claim Networks** | | | | X | X | X | X | X | X | X |
| **4 - Prepare Digital Abtracts** | | | X | | X | X | | | X | |
| **5 - Establish Common Meaning** | | | | | X | | | | X | |
| **6 - Formalize Claims** | | | | X | X | | | | X | |
| **7 - Annotate & Discuss** | | | X | X | X | | X | X | X | |
| **8 - Model Bipolar Claim Networks** | | | X | | X | X | | | X | |
| **9 - Contextualize Model in Documents.** | | X | X | X | X | X | X | X | X | X |

**Table 2: Mapping of Activities to Use Cases for Micropublications.**

The Use Cases in **Figure 2**, with their motivations and uses, are described below. For each one we show a motivation, use, point of implementation, and comments (if any). By "point of implementation" we mean a practical activity already in place in the ecosystem, in which the model could be implemented, within some new functionality that returns value to a user.

1. **Building and Using Citable Claims**

   Motivation. Simkin and Roychodhury used mathematical techniques to show that scientific authors only read approximately 10%-20% of the papers they cite, and many of the rest are evidently copied from reference lists [32, 33]. Greenberg's [3, 4] analysis of the distortion and fabrication of claims in the biomedical literature demonstrates why citable Claims are necessary. In his analysis, it is straightforward to see how citation distortions may contribute to non-reproducible results in a pharmaceutical context, as reported by [8].

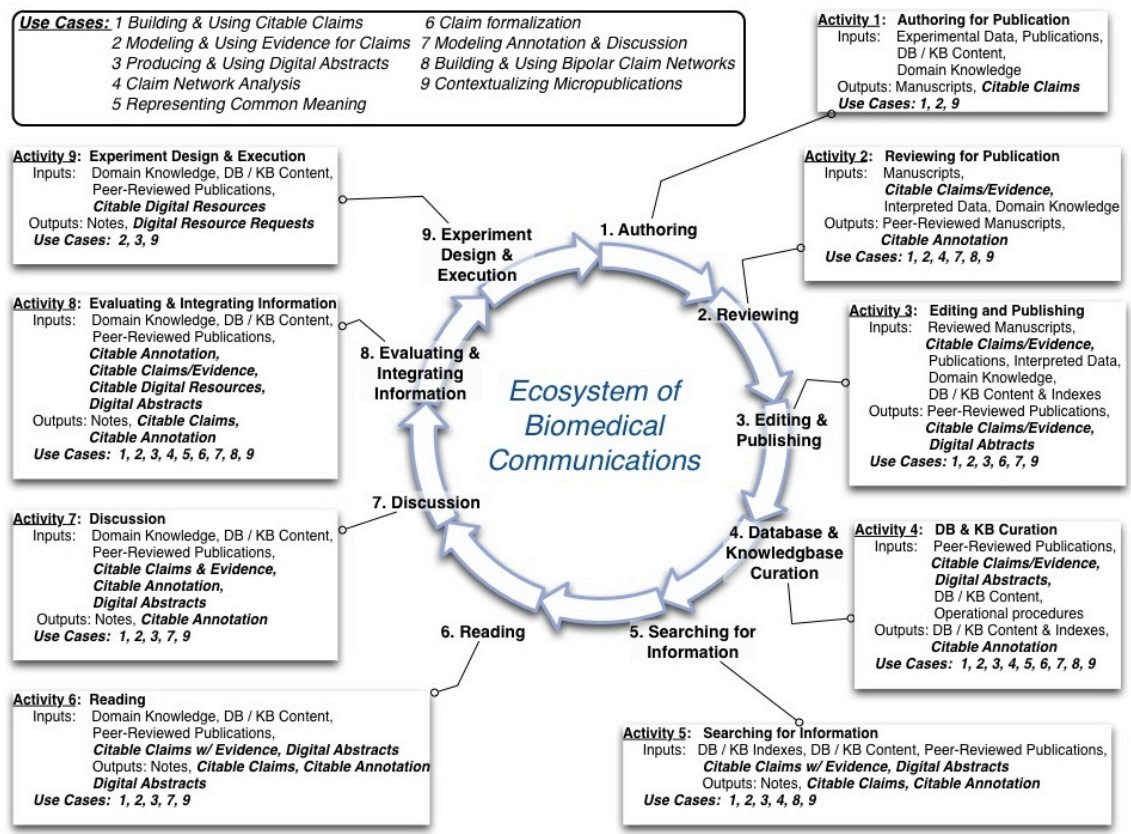

**Figure 1. Activity lifecycle of biomedical communications linked to use cases, activity inputs and activity outputs. The main information content generated, enhanced and re-used in the system is shown. Bolded inputs and outputs represent micropublication-specific content.**

Use: Citable Claims are a specific remedy for citation distortion by allowing ready comparison of what is cited, to what the citation is claimed to assert.

Implementation: Citable Claims may be constructed economically at the point where researchers read and take notes upon, or search for backing for their own assertions in, the domain literature of their field.

Comments: Any scientific statement with an attribution may be formalized as a citable claim using the micropublication model. (Section 4.1)

2. **Modeling Evidence Support for Claims**

<u>Motivation</u>: Evidence is the basis for assessment and validation of claims in biomedical (and scientific) argument. Greenberg specifically showed [3, 4] how citation lineages may not actually resolve to empirical evidence. Claims ultimately must be based on data, and data must be based on reproducible methods. <u>Use</u>: Micropublications may be used to represent experimental evidence supporting Claims, as they can represent non-statement artifacts such as reagents, images and other data. This function of the model has multiple roles. It adds additional value to citable claims by indicating what claims are actually backed by direct evidence, and what this evidence is. It also provides the ability to trace the association of claims in the literature to specific methods and data, and vice versa (Section 4.2).

<u>Implementation</u>: As in Use Case 1, evidence support for claims may be modeled as part of the process of recording bibliographic references. It may also be modeled directly by publishers as supplemental metadata, or by biomedical Web communities as part of a discussion.

3. **Producing a Complete Digital Abstract of a Publication**

<u>Motivation</u>: Digital abstracts would be extremely useful supplemental metadata. They would be particularly useful to enable text mining as argued by Gerstein et al. [36].

<u>Use</u>: A micropublication could be used to formalize the supporting attribution, central claims, references to literature, scientific data, materials & methods, annotations, comments, and formalizations, of scientific communications.

<u>Implementation</u>: They could be provided by publishers or by value-add third parties, or created as part of personal or institutional knowledge bases. Mashing up digital abstracts can be done by third-party applications, and would be one way to deal with intractable publication volumes, by properly summarizing them in a reliable, computable way.

<u>Comments</u>: To enable complete digital abstracting, we define a system of classes for representing biological objects such as reagents, software, datasets and method descriptions, which are not statements in natural language or triples, but are important in documenting the foundational evidence for biomedical claims and arguments, and in making biomedical methods reusable (Section 4.3).

4. **Claim Network Analysis of Publication Sets**

<u>Motivation</u>: As previously noted, it has been shown that the biomedical literature contains a very significant proportion of non-reproducible results. These can be made even more problematic as they are repeatedly cited and transformed in claim networks.

<u>Use</u>: Claim network analysis can be used to determine the origin of, and compare evidence for, individual and contrasting claims in the literature. Particularly when experiments are being designed based on putative findings of a body of prior research; it seems critical to be able to fully assess the entire background of a set of assertions.

<u>Implementation</u>: Micropublications once instantiated, embody individual arguments, which in turn may be composed by resolution of references. This allows us to create extended graphs showing the basis in evidence for claims in the literature, even when they are deeply buried in chains of citations. Claim Lineages are chains of citing/cited Claims. Lineage visualization is proposed as a tool for readers and reviewers (Section 4.4)

5. **Representing Common Meaning using Similarity Groups**

Motivation**:** Resolution of references often entails finding a Claim in a cited document, which is similar to the claim formulated in the citing document. Further, parallel claims, of equivalent or near-equivalent meaning, may arise from different lines of research, without resolution to a common progenitor study.

Use**:** Using *Similarity groups*, a set of Claims may be defined as having "sufficient" closeness in meaning to a representative exemplar, or *holotype* Claim. Their purpose is to allow normalization of diverse sets of statements with essentially the same meaning in the literature, without combinatorial explosion. We term the members of a given equivalence group, *"equivalents"* of one another.

Implementation**:** Holotypes may be defined (a) when a backing statement for a claim is defined, by choosing one or the other as the holotype, or by defining a new "annotator's version" as a holotype; (b) in a similar way, when a text similarity search on the library of claims detects similar statements in separate claim lineages.

Comments**:** The *similog-holotype model* is an empirically based model that allows similar Claims to be normalized to a common natural language representation, without dropping necessary qualifiers and hedging. *Translations* of Claims to formal or other natural languages may also be considered similogs to the translated original, based on (sufficient) equivalence of meaning.

6. **Claim Formalization with Attribution**

   Motivation: Various applications in computing require translation of natural-language claims in the biomedical literature, to statements in a formal vocabulary. Biological Expression Language (BEL) [18] and Attempto Controlled English (ACE) [77-79] are examples of claim formalization vocabularies, as are nanopublications.

   Use: Ideally one would like to be able to trace formalized claims back to their foundational evidence in the literature just as one does with natural language claims. The micropublications model supports formalization of claims. (Section 4.6)

   Implementation: At the point a formalized claim is created (modeled) from a base statement in the literature, the creating application may capture its supporting statement using the micropublications model. For example, in the current BEL software, instead of capturing only the Pubmed ID of the publication from which a BEL statement is derived, one might readily capture the backing statement as well, as a micropublication.

   Comment: Remember that the minimal form of a micropublication is a simple statement, with its attribution, and the attribution of its encapsulating micropublication.

7. **Modeling Annotation and Discussion**

   Motivation: Annotation and discussion of scientific literature is increasingly conducted on the Web.

   Use: Scientific claims and evidence may be annotated in personal or institutional knowledge bases, and may be discussed online in specialized Web portals or communities. Modeling these texts as micropublications, with their backing statements and evidence from the literature, allows them to be exchanged freely between applications in a standard format.

   Implementation: This may be done by Web or other applications at the time the publications are annotated or discussion is captured.

   Comments: See Section 4.7.

8. **Building and Using Bipolar Claim-Evidence Networks**

<u>Motivation</u>: Scientific discourse often involves disagreement on the correct interpretation or theoretical model for existing evidence. It is important to know where gaps or disagreements exist because these naturally suggest areas for further research.   Support/attack relationships exist in the literature for alternative interpretations, hypotheses and models of biomedical function, structure, disease etiology, pathology, agent toxicity, therapeutic action, etc.

<u>Use:</u> We provide an abstract logic representation compatible with much of the current AI literature on argumentation, as well as a description logic presentation modeled in OWL.

<u>Implementation:</u> Where groups or individuals systematically collect statements and evidence on scientific topics, this may be implemented as a useful pattern.

<u>Comments:</u> We believe this approach could be of particular value in drug discovery and development activities. (Section 4.8)

**9. Contextualizing Micropublications**

<u>Motivation</u>: Micropublications may be applied as annotation to scientific documents, including other micropublications. We use an annotation ontology such as AO [56, 57] or OAM [80] to associate micropublication class instances with specific content segments in Web documents, and to record the annotation attribution.

<u>Use</u>: Contextualization is important for the creator of annotations, because it shows them in context.   This is of equal importance for the consumer of annotations.

<u>Implementation:</u> Implementation can be within the annotating application.

<u>Comments:</u> See Section 6.1. We believe micropublications will most commonly be created as annotations.

The most important thing to note about the activities constituting the use cases, is that all of them involve assembling, justifying, critiquing, or representing some form or elements of scientific argumentation, including all the support for the argumentation, i.e including the empirical evidence. Thus, very few of these use cases can be met adequately by purely statement-based models.

After presenting our model of argument in abstract form, and comparing it with a statement-based approach, we continue our analysis to show how this model would be deployed in examples illustrating each of the above use cases, and then give a concrete version of the model as an OWL vocabulary [81], with RDF [82, 83] instantiations of several use case examples. The OWL Micropublications vocabulary can be applied as semantic metadata to characterize, in a conceptually uniform way: the text of scientific documents; the relationships of argumentation elements of the text within documents; annotations upon those documents; and formalizations of scientific statements within them.

## 3    The Micropublications Model

The Micropublications model is a framework which accomodates a spectrum of complexity, from minimal to maximal representations.  It can ingest the simplest forms and give room for stepwise elaboration, consistent with the incremental distributed value chain in which we are trying to embed it. The minimal representation is a single identified statement, where attribution is attached to both the statement and the identification.  The maximal representation may be as complex as an entire knowledge base.

To introduce this model, we first describe some modelling considerations, and then outline our semantic and mathematical models of argument.   Next we illustrate examples of how the model may be applied to each of the Use Cases from the preceding analysis, across the cycle of activities depicted in Figure 1. Lastly we present the base classes and predicates of the full model, and some accompanying Rules.

It is worth stressing that despite the richness of this model, using it does not by any means require deploying all of its concepts for any particular scenario.

### 3.1 Modelling Considerations

Constraints on this model are imposed not only by the Use Cases as they relate to biomedical scientists, but also by other work in the field of argumentation models (as reviewed in [53]), which we would like to be able to reuse where possible. We would like to model both the internal structure of arguments, to digitially summarize publications in a useful way; and the interargumentation structure, so as also to model relations between arguments considered without regard to their internal structure. Also, we want to enable construction of claim networks similar to that described by Greenberg [3], which are principally based on support relationships, but may also have a significant challenge or attack component.

While Toulmin [44] and Verheij's [45, 46] approaches might suggest having different relationship types between entities in the model, and the use of *backing* and *warrant* as classes; we avoid that, because these concepts become relativized across a large network: one publication's *backing* is another's *warrant*.

This relativization suggests a graph structure, which is also compatible with work in unipolar [47] and bipolar [49] argumentation frameworks and in claim-evidence networks à la *Hyper* [84] or SWAN [20]. Using a graph model requires common connective properties to allow transitive closure. So for example, the relation between data and its interpretation in a textual statement is called *support*, as is the relationship between a statement and the reference cited to justify it. In another context, these might be modelled as disparate properties, say as "interpretation" and "citation".

### 3.2 Logical Formalization

### 3.2.1 Introducing the Model: How Micropublications Represent Argument

Micropublications represent scientific arguments. The goal of an argument is to induce belief [44]. An argument (therefore a Micropublication) argues a principal *claim*, with statements and/or evidence deployed to support it. Its support may also include contrary statements or evidence; and/or the claim may dispute claims made by other arguments. These are called *challenges* in our model, *rebuttal* by Toulmin [44-46], and *attacks* in the artifical intelligence literature on argumentation frameworks (see e.g. [47, 49, 54, 85].

The minimal form of an argument in our model is a statement supported by its attribution. If the source of the statement is trusted, that may be enough to induce belief. Aristotle called this aspect of rhetoric *ethos*, the character and reputation of the speaker [86]. **Figure 3** shows this minimal form of Micropublication.

The support of a Micropublication's claim is structured as a graph. Unlike the standard Toulmin-Verheij model, which only deals with statements, scientific argument must ultimately support statements with empirical evidence, consisting of

- data in the form of tables, images, etc.; and
- descriptions of the reproducible methods by which this data was obtained; which may include drawings, photographs, etc.

Scientific argument must also situate its claim in the context of previous work in the domain, of which it takes account – as additional support, or as error to be challenged and disproven. This context is deployed as paraphrases of other published findings (claims), qualified by a citation of the work from which they were paraphrased. Toulmin calls these paraphrases "warrants" (as in, "warrants for belief"), and the work indicated by a citation is called the "backing", which would be consulted to validate the warrant.

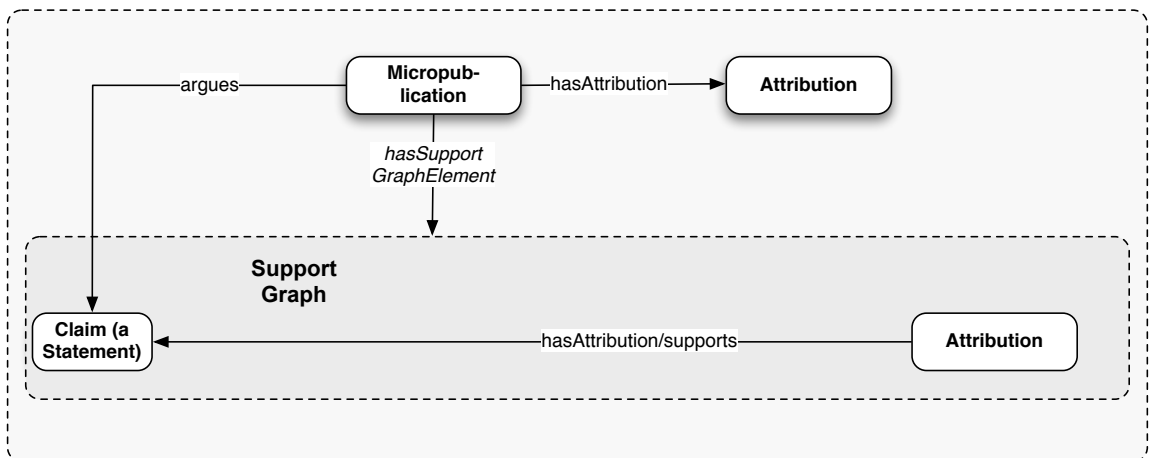

**Figure 3. Minimal form of a Micropublication: formalizing a Statement and its Attribution.**

As we are interested in constructing claim networks, it should be clear that in a network, warrant and backing are relative terms. Furthermore, to contruct such a network, we will need to have backing which resides in another work, available in the form of a single statement, not the entire work. While a citation of an entire article may be acceptable as a temporary measure, reflecting pragmatic boundaries, ultimately we wish to have the full claim network at hand. This sets us up to be able to transitively close the network. To do so we use a *supports* relationship between warrant and backing.

Defining this relationship consistently across the model – whether we are dealing with supporting statements, data, or methods – also allows us to bridge the gap between internal argument structure, and inter-argument structure.

We call any element of an argument, a *Representation*, a class whose subclasses include *Sentence*, *Statement*, *Claim*, *Data*, and *Method*. *Sentences* need not be syntactically complete – they may consist of a phrase, single word, or single meaningful symbol (e.g. "We hypothesize that", "Often", "¬"). Declarative *Sentences* are *Statements*. Sentences which qualify a Statement are *Qualifiers*. The principal *Statement* in an argument is called a *Claim*. Statements may be supported by other Statements, or by *Data*. Data in turn may be supported by *Method*, i.e. a description of how the Data was obtained, in the form of a re-usable recipe (or recipe component). A *Procedure* or a *Material* is a Method.

**Figure 4** shows a simplified model of an argument using this approach, and without deeper semantic characterization of its elements. The Claim is supported here by both a Statement paraphrasing another finding in the literature, and by Data. The paraphrase is supported by a Reference to the work in which we are supposed to be able to find its source. The Data is supported by its Method. Both the Micropublication itself, and the argumentation it formalizes, have Attribution. All *elementsOf* of the Micropublication which support its Claim, are in its *SupportGraph*.

Later we will examine various forms of argument formalized as Micropublications, using a closely related set of examples taken from the literature on Alzheimer Disease, for each Use Case.

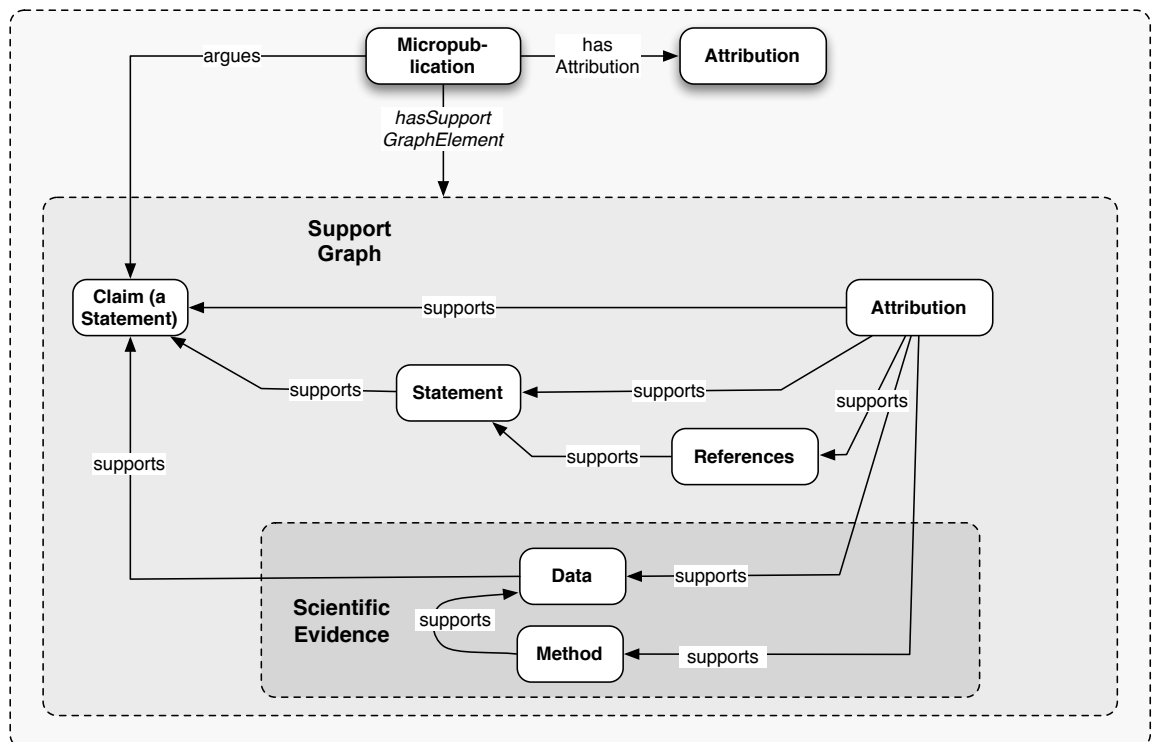

**Figure 4. A Micropublication supported by a Statement referenced to the domain literature; empirical Data; and a reusable Method.**

### 3.2.2 Outline Semantic Representation of the Model

The basic outlines of our model are given here more systematically. An OWL model corresponding to these definitions is available at http://purl.org/mp, and detailed definitions are provided in the Supplementary Material.

1.  Entities, Agents, Artifacts, Activities and Representations.

    a.  *Entities* are things which may be discussed, real or imaginary.
    b.  An *Agent* is an Entity that makes, modifies, consumes or uses an *Artifact*.
    c.  A *Person* or an *Organization* is an Agent.
    d.  An *Artifact* is an Entity produced, modified, consumed or used by an *Agent*.
    e.  An *Activity* is a process by an Artifact is produced, modified, consumed or used.
    f.  A *Representation* is an Artifact which *represents* something.
    g.  A Representation may be a *Sentence*, *Data*, *Method*, *Micropublication*, *Attribution, or ArticleText*.

2.  *Supports* and *challenges*.

    a.  The *supports* property is a transitive relation between Representations.

    b. The *challenges property* is inferred when a Representation either *directlyChallenges* another, or *indirectlyChallenges* it by undercutting (*directlyChallenges*) a Representation which *supports* it.

3. Sentences, Statements, Claims and Qualifiers.

    a. A *Sentence* is a well-formed series of symbols intended to convey meaning.

    b. A *Statement* is a declarative *Sentence*.

    c. A *Claim* is the single principal Statement *arguedBy* a Micropublication.

    d. A *Qualifier* is a Sentence, which may modify a Statement. *References* and *SemanticQualifiers* (tags) are two varieties of Qualifier.

4. Data, Method and Material.

    a. *Data, Method* and Material are kinds of Representation.

    b. If Data *supports* a Statement, that Statement is *supportedByData*. Data may be *supportedByMethod* if a Method *supports* it.

    c. A *Method* is a reusable recipe showing how the Data were obtained; it specifies an Activity, and may refer to some Material as a component of the recipe. A Material *supports* any Method of which it is a component.

5. Micropublications

    a. A *Micropublication* is a set of *Representations*, having *supports* and/or *challenges* relations to one another and potentially to those which are an *elementOf* other Micropublications.

    b. A *Representation* is defined as an *elementOf* a Micropublication if that Micropublication either *asserts* or *quotes* it.

        i. A Representation *assertedBy* a Micropublication is originally instantiated by that Micropublication.

        ii. A Representation *quotedBy* a Micropublication is referred to by that Micropublication, after first being instantiated by another Micropublication.

        iii. The asserts and quotes notions are very simple extensions of concepts from Carroll et al. and Bizer's work on provenance and trust [87, 88].

    c. *Claims*

        i. A *Claim* is the principal statement *arguedBy* a Micropublication.

        ii. The *supports* relationships amongst the Representations in a single Micropublications are structured as a directed acyclic graph (DAG), whose root is the Micropublication's *Claim*.

    d. *Attributions*

        i. The minimal level of support for any Statement is its *Attribution* to some *Agent*.

    e. Support and Challenge Graphs

        i. Representations related to the Claim of a Micropublication by the *supports* property, and which are *elementsOf* that Micropublication, constitute its Support Graph, being related to the Micropublication by the property *hasSupportGraphElement*.

        ii. The property *hasChallengeGraphElement* works similarly.

Figure 5 shows the major classes and relationships in the model.

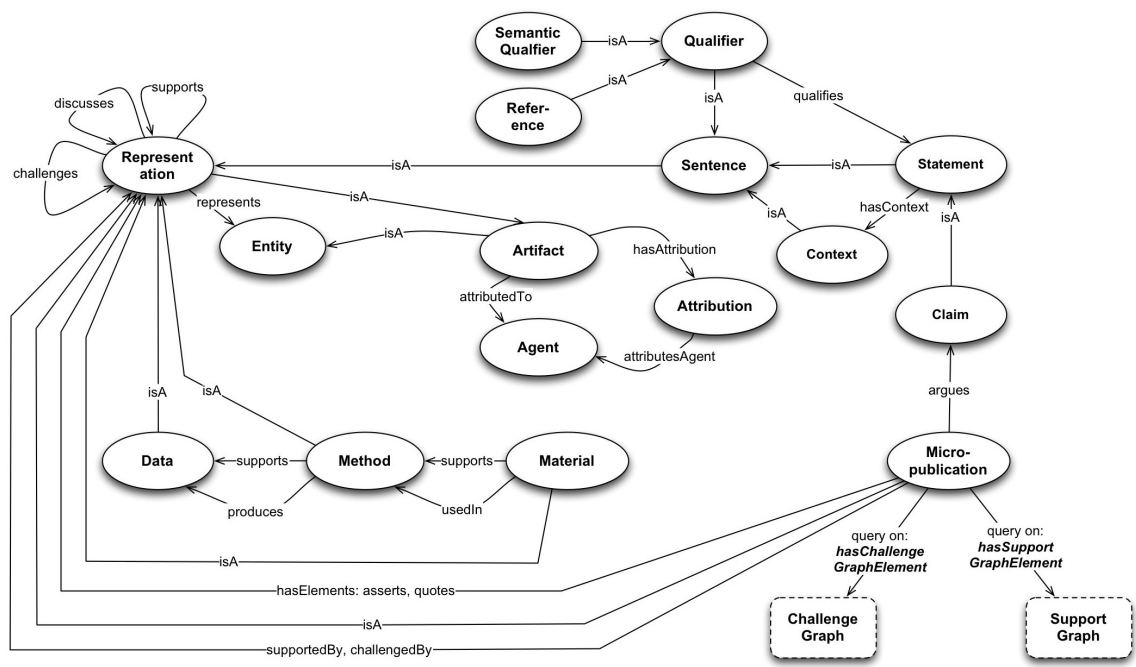

**Figure 5. Major classes and relationships in the model. A** *Claim* **is the main** *Statement* **argued by a** *Micropublication*. **A Statement is a truth-bearing** *Sentence*, **which may be variously** *qualifiedBy* **some** *Qualifier*. **A Sentence is a well-formed sequence of symbols, intended to convey meaning; and is not necessarily either complete or truth-bearing. A Micropublication** *hasElements* **consisting of those** *Representations* **it** *asserts* **or** *quotes*. **A Representation** *supports* **or** *challenges* **other Representations. The supporting Representations which are** *elementOf* **the Micropublication will be in its** *SupportGraph*; **challenging elements, will be in the** *ChallengeGraph*. **Dashed-line boundaries indicate graphs instantiated by query.**

The class Artifact has, as previously noted, a series of subclasses allowing us to deal relatively homogeneously with experimental methods, materials, data, and language artifacts such as statements.

All Statements, as Artifacts, should have an Attribution. The Attribution of a Statement is therefore a part of its SupportGraph. The simplest Micropublication would be an instance of the Micropublication class, with its Attribution; arguing a Claim, supported by the Claim's Attribution.

In scientific argumentation, a publication in *Science*, *Nature*, etc. by an eminent scientist, concerning his or her own principal area of expertise, is typically given more weight than an article in a third-tier journal, or a blog post or tweet from an undergraduate with limited expertise. This is why Attribution is part of the SupportGraph of an Argument. However, Attribution alone is weak support. The critical element of support in science is empirical Scientific Evidence.

Complete class and predicate definitions for the model are given in section 3.2.4.

**Figure 6** shows a Micropublication based upon text from Spilman et al. 2010 [27]. It includes an example instantiation of *SemanticQualifiers*.

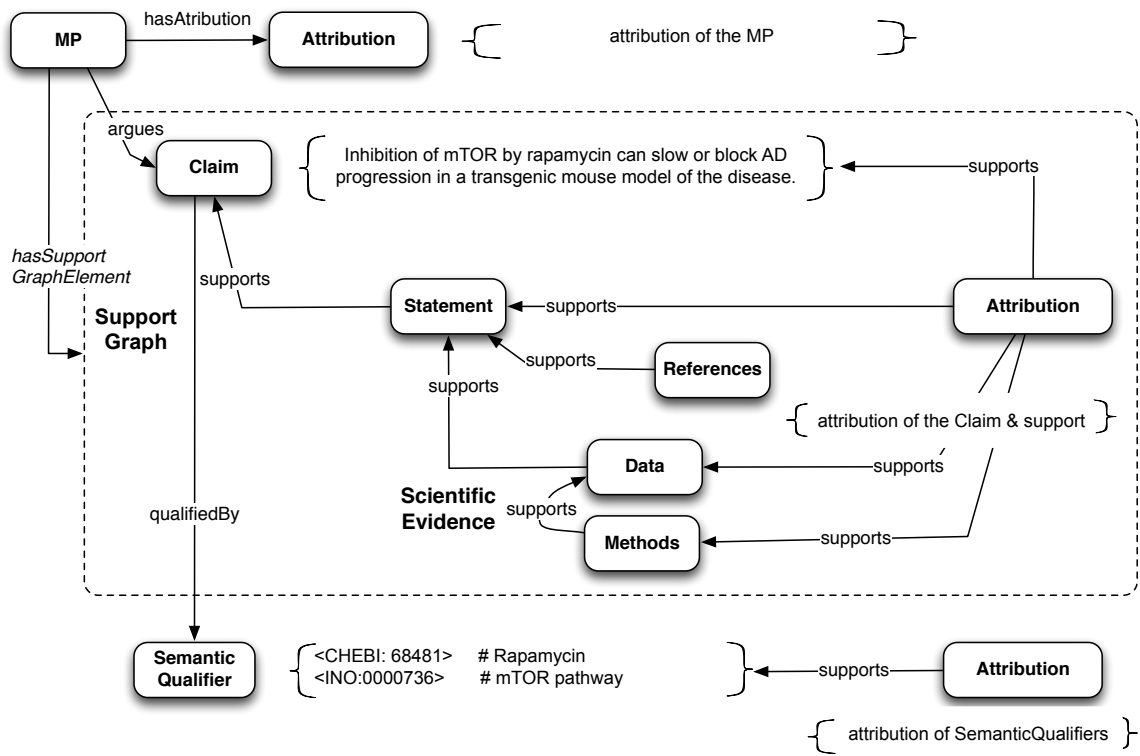

**Figure 6. Representation of statements and evidence in a micropublication format, based on text from Spilman et al. 2010 [27]. Micropublications show the graph of support for a Statement, including the author's Data and Methods, constituted by a set of *supports* relations in the SupportGraph. The Claim is qualifiedBy SemanticQualifiers <CHEBI:68481< and <INO:0000736>, which are abbreviations for the respective fully-qualified URIs.**

### 3.2.3   Abstract Mathematical Representation of the Model

Let $a$ denote the text of an argument – which in most of our examples, will be sections of a scientific article.  It may include images and other data as well as text.

Let  MP_a denote the corresponding formalization of $a$ as a Micropublication.

Then MP_a = ⟨A_mpa, c, A_c, Φ, R ⟩ is the representation of $a$ as a **Micropublication** or formalized argument, where

- $a$ ::= an argument text, represented in a document;
- MP_a ::= a micropublication, or formalized argument structure, defined on $a$;
- A_mpa ::= the Attribution of this formalization of $a$ as a micropublication;
- c $\in$ Φ ::= a single Statement, being the principal Claim of $a$;
- A_c $\in$ Φ ::= the Attribution of the Claim c; and
- Φ  is  a  finite  non-empty  set  of  Representations  which  are  elements  of  the Micropublication;
- R⊆ Φ✕Φ;  R being a nonempty set of *supports* and *challenges*  relations, $r(\Phi_i,\Phi_j)$ | R is a strict partial order over Φ and its greatest element is c, the principal Claim of the argument in $a$.

4    Case Studies and Design Patterns

Here we present a series of case study examples showing the application of the model to a publication, the structure of its argument, backing in the literature, and concrete scientific evidence. These examples begin with the use case of a single referenced statement and proceed to more complex cases.  The models in these case studies constitute a set of design patterns.

In each Example we abstract an argument from the biomedical literature.  After illustrating the abstract form it assumes as a Micropublication, we diagram its structure, and describe a use case. In most of the following diagrams, for simplicity, we do not show the Attribution for the Claim and the SupportGraph elements.

RDF examples for several of these use cases are provided in the Supplementary Material.

### 4.1    Example 1: Citable Claim with Supporting Reference and Attribution

*Use Case*: The base use case for Example 1 is constructing directly *citable Claims*.

*Model:* The simplest form of a Micropublication represents a Statement with its supporting reference Attribution.  In this form it is similar to a nanopublication with text representation of the claim substituted for triples. The argument in this and many subsequent examples is taken from  Spilman et al. 2010 [27].

Let the the argument $a_1$ = "Rapamycin is an inhibitor of the mTOR pathway (Harrison et al., 2009)." Then for MP1, the formalized argument (micropublication) derived from $a_1$, we have

- o  the Claim (C1) is that "Rapamycin is an inhibitor of the mTOR pathway";
- o  the Attribution (A_C1) of this claim is to the Agent *PSpilman*;
- o  the Claim is *qualifiedBy* the *SemanticQualifiers* CHEBI: 9168 and INO_0000736;
- o  the SupportGraph for the claim is the set {supports(Ref5,C1), supports(A_C1, C1) }.

**Figure 7** illustrates the structure of Example 1.

*A note on Toulmin-Verheij terminology*: The Claim in **Figure 7** would be a *Warrant*, in Toulmin-Verheij terminology, and the Harrison et al. reference is to the *Backing* of this Warrant. The Warrant supports belief in the asserted Claim, as either a key supplement to, or in this case, in lieu of, direct evidence (i.e. scientific data) presented in the argument. This micropublication consists only of literature references. We do not assume a cited reference is necessarily valid – it is a representation which may or may not resolve to a real document. If it can be so resolved, we connect it with a further *supports/supportedby* relationship, to some unique identifer of that document. If the document has Micropublication metadata associated with it, we can then potentially further resolve the reference, to a specific Statement within the document.

*Use Case Detail*: The base use case for Example 1 is constructing *libraries of citable Claims*.

Greenberg's analysis of citation distortion [3, 4] makes a clear case for citable Claims. The elements of a Claim's SupportGraph are its Warrant for belief, in addition to any empirical supporting evidence directly presented. The Warrant is a purported summary of the Backing, which in Example 1 would be the actual document referred to by the citation [69].  If the Warrant does not adequately represent, distorts, or fabricates its Backing, and this is known, the validity of the Claim it supports is called into question.  Greenberg demonstrated that this happens far more frequently than one would like and is a serious defect in the domain literature (amounting to hundreds of papers) which he reviewed. Because it is far too laborious to check each and every cited document, searching for the relevant Claim, citable Claims are proposed here as a method to dramatically reduce the labor cost of checking a Claims' support. With proper software, identifying such Claims may be automated.

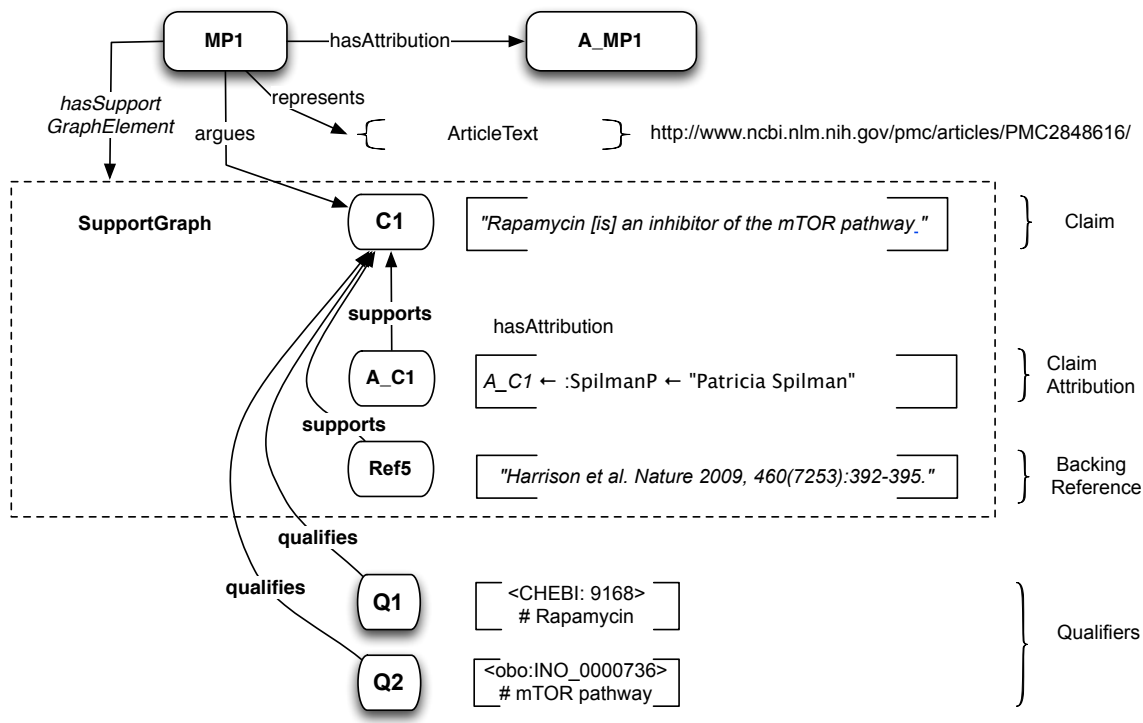

**Figure 7. The argument "Rapamycin is an inhibitor of the mTOR pathway" represented as a micropublication, with semantic qualifiers. This argument is taken from Spilman et al. 2010 [67]. C1 is the Claim; A_C1 is the Attribution of the Claim; Ref5 is the Claim's supporting reference; SG1 is the SupportGraph. At the time the micropublication was extracted, its claim was assigned the Qualifiers Q1 and Q2. Note that the Claim Attribution for C1, represents the attribution of the article in which the text of C1 appears, not the article cited in support of C1.**

Citable Claims could be constructed most economically at the point where researchers read and take notes upon, or search for backing for their own assertions in, the domain literature of their field. Today researchers commonly use bibliographic reference managers such as Zotero, EndNote, etc. to record this aspect of reading and note-taking, but the Claim(s) for which they record a bibliographic reference are not captured. Our model allows them to have a standard sharable representation.

### 4.2 Example 2: Modeling Evidence Support for Claims. Citable Claims with Supporting Data and Reproducible Methods

*Use Case:* The base use case for Example 2 is to enhance *citable Claims* with *supporting Data and reproducible Methods*.

In **Figure 8** we abstract and model a Statement from Spilman et al. 2010 [27], supported by *Scientific Evidence*, i.e., a Representation of *Data*, and the *Methods* (= "materials and methods") by which it was obtained, including pre-observation interventions and observational context.

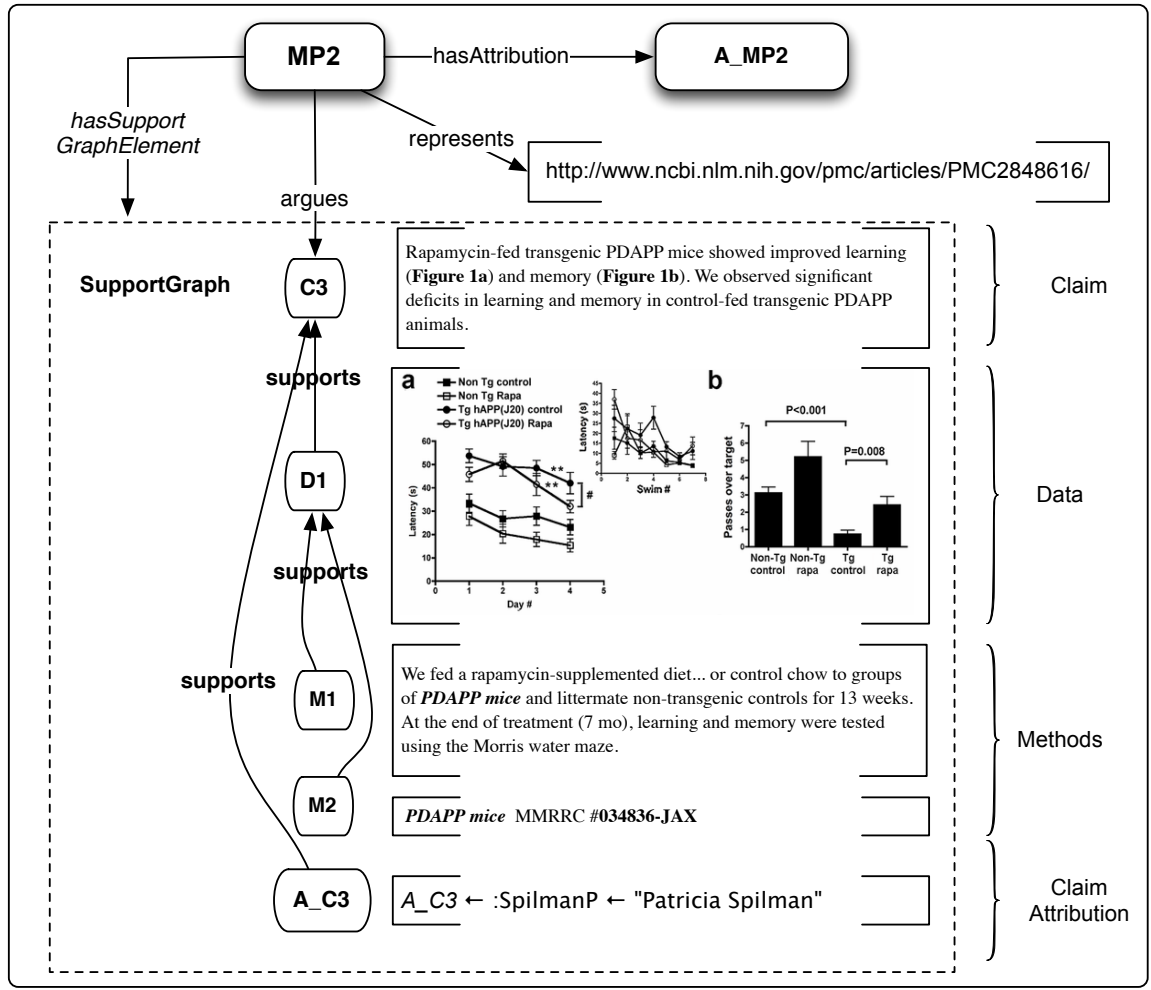

**Figure 8. Scientific Evidence consisting of Data and Methods supporting C3, a Claim from [67]. D1 is a composite image of graphs produced from the primary data. The Methods M1 (a procedure) and M2 (a transgenic mouse strain), both support D1.**

Let $a_2 =$

- the Claim indicated by C1 in Figure 5 and
- the supporting Evidence indicated by D1, M1 and M2 in Figure 5.

Then for the formalized argument or micropublication MP2, we have

- the Claim is C3;
- the Attribution of this claim is A_C3, which in turn is related to the Agent *PSpilman* by the *attributionOfAgent* property; *PSpilman* is "Patricia Spilman";
- the Support Graph is a set of relations {supports(A_C3,C3), supports(D1,C3), supports(M1, D1), supports(M2, D1)}.

*Use Case Details*: Citable Claims with supporting data and resources give us the capability of showing the material basis for Claims based on original research in a given work.  Also, in the forward-looking case of citable archived data and resources, the claims associated with them might be more easily determined. Citable claims in a personal or institutional "library" or database may "index" the associated data and methods more readily than other forms of metadata. Lastly, methods and materials later found to be flawed might be easily traced to claims based upon them.

### 4.3    Example 3: Computable Digital Summary of a Publication

*Use Case*: The base use case for Example 3 is digital abstracting of a biomedical article in computable form, based on citable Claims with all supporting Statements, Data and Methods.

This more complex form of Micropublication summarizes the principal Claim of an article and the evidence supporting it, again based on [27], and is illustrated in **Figure 9.**

Let $a_3$ =
- o    the Claim indicated by C3 above;
- o    the supporting Statements indicated by S1, S2 and S3; and
- o    the supporting Evidence indicated by D1, M1 and M2.

Then for MP3 its micropublication we have
- o    the Claim is C3, "Inhibition of mTOR by Rapamycin can slow or block AD progression in a transgenic mouse model of the disease.";
- o    A_C3 is the Attribution of C3, with *PSpilman* as its object via *attributionForAgent*, as in the previous example;
- o    the SupportGraph is a set of *supports* relations on {A_C3, S1, S2, S3, D1, M1, M2, Ref5, Ref9, Ref10}.

### 4.4    Example 4: Claim Network Analysis Across Publications

*Use Case*: The base use case for Example 4 is Claim network construction and analysis across publications.  This supports the need of researchers to record and see clearly the actual statements and evidence intended to be referenced in a cited publication, as opposed to taking on faith the citation of an entire document treated as a "black box".

Claim C3 in **Figure 10**, relies on three logical elements:
1.  Rapamycin inhibits mTOR (taken from the literature);
2.  PDAPP mice are a good (Spilman actually says "established") model of human Alzheimer Disease pathology (also from the literature); and
3.  Experimental evidence that PDAPP mice fed Rapamycin over time regain some measure of cognitive health.

A flaw in any of these elements undercuts Claim C3, which is the essential argument of the publication. So it is worthwhile in to examine, in addition to the experimental evidence, whether the two supporting statements from the literature are well-grounded.  Careful readers do this for important points – or may have the relevant citations already memorized.  Implementation of the present use case allows this information from the literature to be modeled, retained and shared.

Suppose that for the supporting literature references in Example 3, [69] and [71], we had also created micropublication annotations over their target documents.  It is not unrealistic, given some appropriate software to be used by researchers when looking up the desired claim in a work to be cited, to record it as a citable Claim.

Resolving the Claims in Spilman's argument to their support in the backing references, and connecting the citable Claims, gives the graphs shown in and **Figures 10 and 11**.

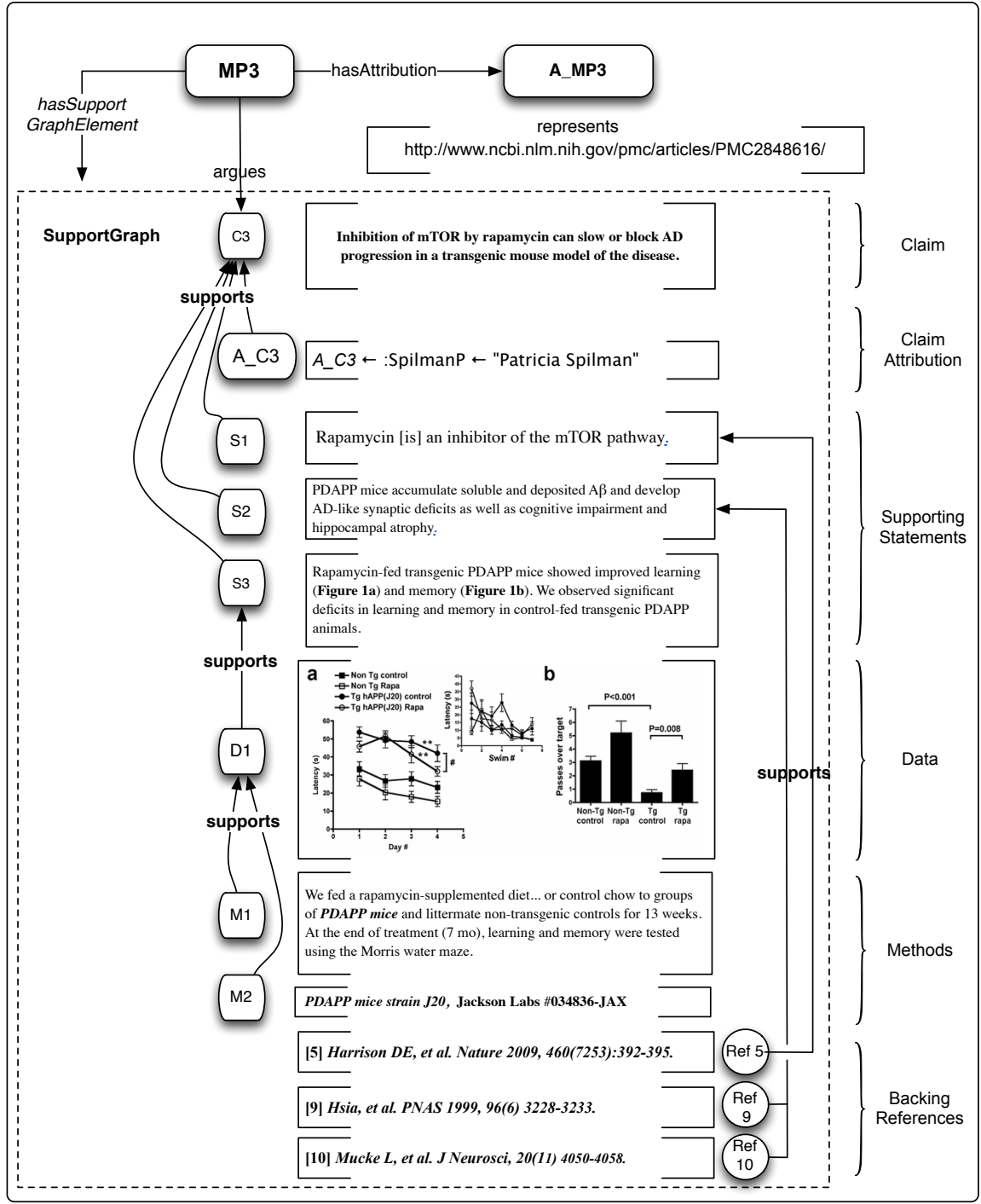

**Figure 9. Micropublication for the principal Claim of Spilman et al. 2010 [67]. Claim C3 from the abstract is supported by Statements S1 and S2, with support in the literature (Ref5, Ref9 and Ref10); and S3, which interprets the Data in D1. D1 in turn is supported by Methods M1 and M2. All elements of the Support Graph have Attribution A_C3; for simplicity we show only the relationship *supports*(A_C3, C3) in this figure.**

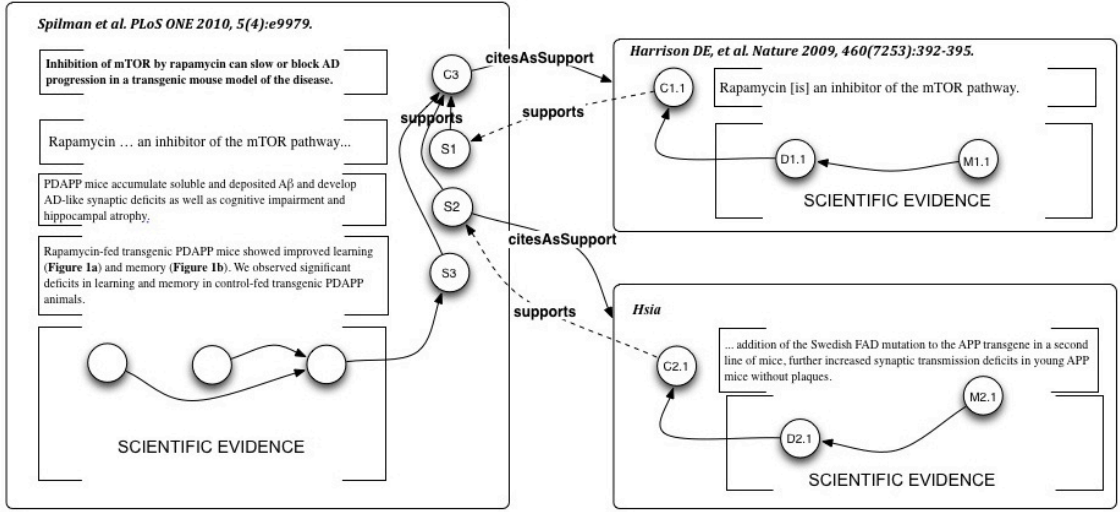

**Figure 10. Transition from document-level backing references, to claim-level backing references, to construct a citation network.**

The citing Statements in [27] are now connected directly and transparently to their backing in other publications at the level of Claims. The backing Claims in the two cited publications are in turn connected directly to supporting primary scientific evidence, which may be inspected.

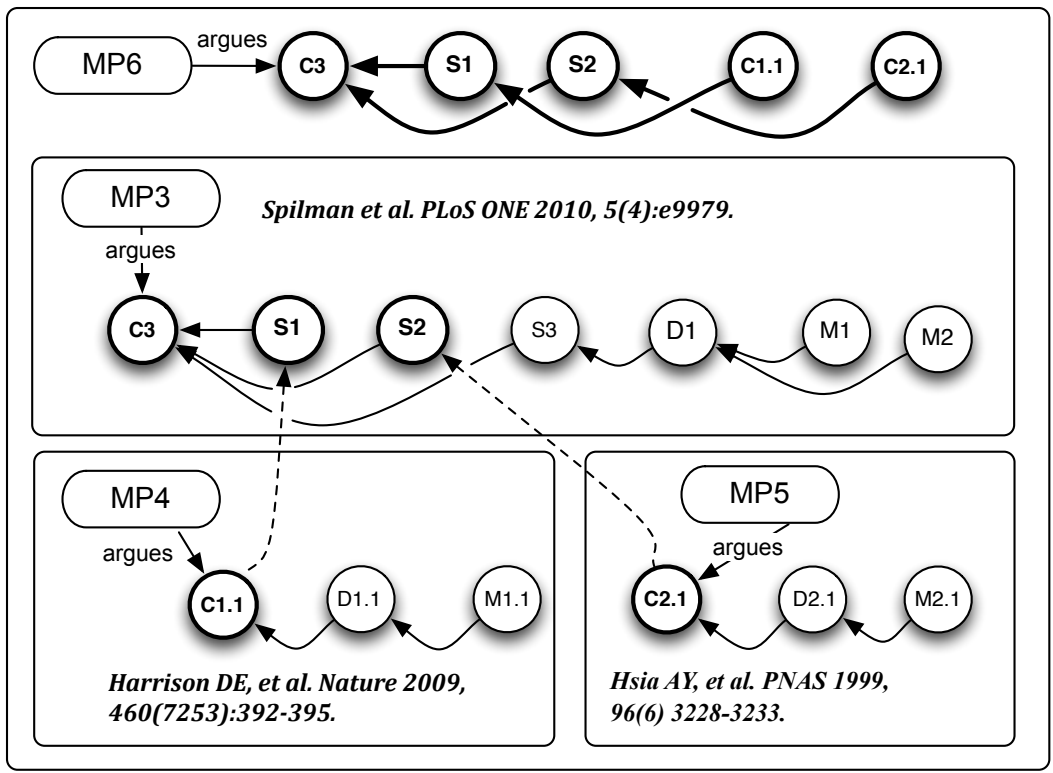

**Figure 11. Connected support relations of three arguments give a Claim network across three publications. Micropublication MP6 asserts the new information that C1.1 is support for S1; and C2.1 is the support for S2. The dotted lines connecting elements of MP3, MP4 and MP5 are given by the support relations in MP6.**

Spilman's Statement S1 ( "Rapamycin … an inhibitor of the mTOR pathway…") cites [69] in support; inspection of the cited article shows the corresponding Claim in Harrison et al. ("Rapamycin … an inhibitor of the mTOR pathway…"), shown as C1.1 in **Figures 10 and 11**.

Spilman's Statement S2 ("PDAPP mice accumulate soluble and deposited Aβ and develop AD-like synaptic deficits as well as cognitive impairment and hippocampal atrophy cites [71], shown as C2.1 ("…addition of the Swedish FAD mutation to the APP transgene in a second line of mice, further increased synaptic transmission deficits in young APP mice…") in **Figures 10 and 11**.

Close inspection of the Backing for S2 shows how valuable it may be to clarify this at a Claim-to-Claim level; and to cite actual reagent catalog numbers as Methods (see [58] for discussion). The "second line of mice" referred to here turns out to be the J6 line of PDAPP mice.

Both the J6 and the J20 lines are from the same lab. But Spilman et al. used the J20 line, not the J6. The J20 line, Jackson Labs #034836, is barely mentioned in Hsia et al., though both come from the Mucke laboratory. This is perhaps why earlier in the article, the authors simply referred to "PDAPP mice" in general in asserting that they were an "established model" of AD.

Why was a different line used that the one discussed in the actual backing of the reference Hsia et al.? Perhaps it is worth investigating…perhaps the J20 line was readily available, but the J6 line was actually documented. What this points to is that the lines are not sufficiently documented in this article, to the point that authors cite them in a way which seems deliberately obscure. We will ask further questions about these transgenic mouse lines later. But already, the model has helped to show the discrepancy between methods citations, and the actual methods used.

In resolving support across Micropublications, we establish a Claims network. However we must ensure we retain localization of responsibility, for what is originally presented in the merged graphs, and what is "imported" from elsewhere. This is done using the *asserts* and *quotes* predicates. In the example, MP6 *quotes* C3, S1, S2, C1.1 and C2.1. It *asserts* the (new) support relationships between C1.1 / S1, and C2.1 / S2.

*Abstract model of the Claim about Rapamycin:*

Let $a_4$ =
- o      Claim C1.1 from **Figure 10**, and its supporting scientific evidence.

Then for MP4, its micropublication, we have
- o      the Claim is C1.1, "Rapamycin … an inhibitor of the mTOR pathway…", derived from "Harrison et al. 2009";
- o      the SupportGraph is the set of support relations {supports(D1.1,C1.1), supports(M1.1,D1.1)};

giving us a micropublication model for the Rapamycin Claim from [69].

*Abstract model of the Claim about PDAPP mice:*

Let $a_5$ =
- o      Claim C2.1 from **Figure 10**, and its supporting scientific evidence.

Then for MP5 we have
- o      the Claim is C2.1, "... addition of the Swedish FAD mutation to the APP transgene in a second line of mice, further increased synaptic transmission deficits in young APP mice without plaques.", derived from "Hsia et al. 1999"; and
- o      the SupportGraph is the set of support relations on {C2.1, D2.1, M2.1};

giving us a micropublication model for the PDAPP mice Claim from [71].

*Abstract model of the Spilman et al. Statements, with Backing resolved:*

Let $a_6$ =
- o     Claim C3 from **Figure 10**; and
- o     the supporting Statements indicated by S1, S2 and S3.

Then for mp($a_6$) we have

- o     the Claim is C3, "Inhibition of mTOR by Rapamycin can slow or block AD progression in a transgenic mouse model of the disease.", from Spilman et al. 2010;
- o     the SupportGraph is expanded and now adds connections to C1.1 and C2.1.

We call the graphs C1.1→S1 and C2.1→S2, Claim Lineages, by analogy with biological lineages.

### 4.5   Example 5: Representing Statements with Similar or Identical Meaning: Similarity Groups and Holotypes

In it is easy to see that C3 and C1.1 mean the same thing. C3 is derived from C1.1 (in Toulmin's terminology, a "Warrant"). We call these Statements *similogs* of one another. Groups of *similar* statements are equivalence classes, defined as having "sufficient" closeness in meaning to a representative exemplar, or *Holotype* Claim.

A Holotype, or representative of the genus, is selected as a matter of convenience and exemplification, to stand for the common meaning of the Statements in a similarity group.

In addition to the C1.1→C1→ Lineage, we have three other publications, [89], [90] and [91], containing Claims C4, C5 and C6, with similarity to C1.1 & C3; C4 is selected as representative of the group. It could very legitimately have been cited as support by C3. The Sabatini paper, source of Claim C6, is one of the three original articles from 1994 published on this interaction and provides extensive primary evidence. It refers to "RAFT1", a synonym for the preferred protein name mTOR, which superseded it in the nomenclature.

The group-of-similogs / holotype approach is empirical. It is based on the notion that scientific communications in form and content are a "literary technology"[a] [29], which mediates collaboration and exchange of knowledge amongst scientists. As a form of technology, Statements, and the concepts they express, evolve and build upon one another through social interaction. This is a practical alternative to the idea that "sentences *express* Statements", originating with Strawson's work [92], which considers Statements as extra-mundane ideal abstractions standing apart from the world of real texts as they are exchanged and discussed [b].

Another question now emerges: identification of statements as being *similogs* is itself an assertion. Your "similarity" may not be my "similarity", depending upon both the particular application (think of this as a kind of "manufacturing tolerance"), or in some cases, professional judgement. Can we model such an assertion as a micropublication? Yes, as shown in **Figure 13.**

In Strawson's perspective, we cannot assign any explicit ontological status, i.e. material existence, to a Statement. It is an extra-mundane abstract "meaning" – but always reduced to "meaning" as expressed in the language of formal logic, which is thus smuggled in via the back door. This would seem to be highly problematic, and we avoid it, because scientific publications for the most part do not represent their findings in this way. We then incur a translation step, which can do violence to the natural language presentation of the actual publication. In our model, Statements are Statements, i.e. declarations, assertions, truth-bearing Sentences, in some meaning-conveying language, and do have explicit ontological status.

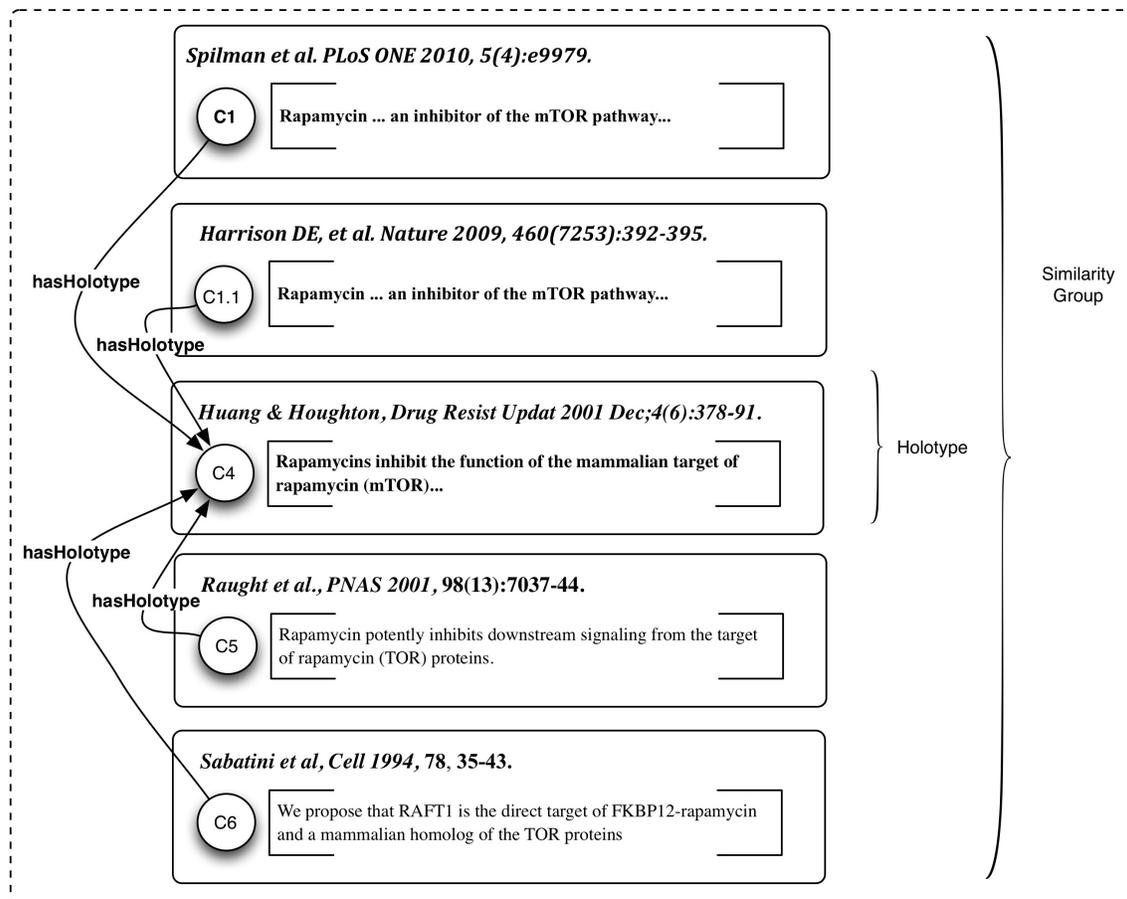

**Figure 12.** *Similarity Group* **with representative (***Holotype***) Claim. The Holotype is from one of three additional publications, outside the Claim Lineage C1➔S1.**

### 4.6    Example 6: Claim Formalization In Biological Expression Language

*Use Case*: Relating a formal-language model of the content (meaning) of a Claim using a controlled vocabulary, directly to its Backing in the literature. It can be useful to translate the content of textual Claims in a scientific publication, into a specialized formal language, for specific computational tasks - for example, doing systems biology.

The Biological Expression Language (BEL) [18] is one example of such a formalism, which is used to construct knowledgebases of molecular interactions by the pharmaceutical industry.  The current software support for BEL associates one or more PubMed identifiers with every BEL statement.  As in the case of ordinary textual Claims citing entire documents, it is then laborious to reconstruct the actual Backing when need arises.  Yet at the original point of extraction of the BEL statement, that Backing was readily available and could have been cited directly given a suitable model and associated software.

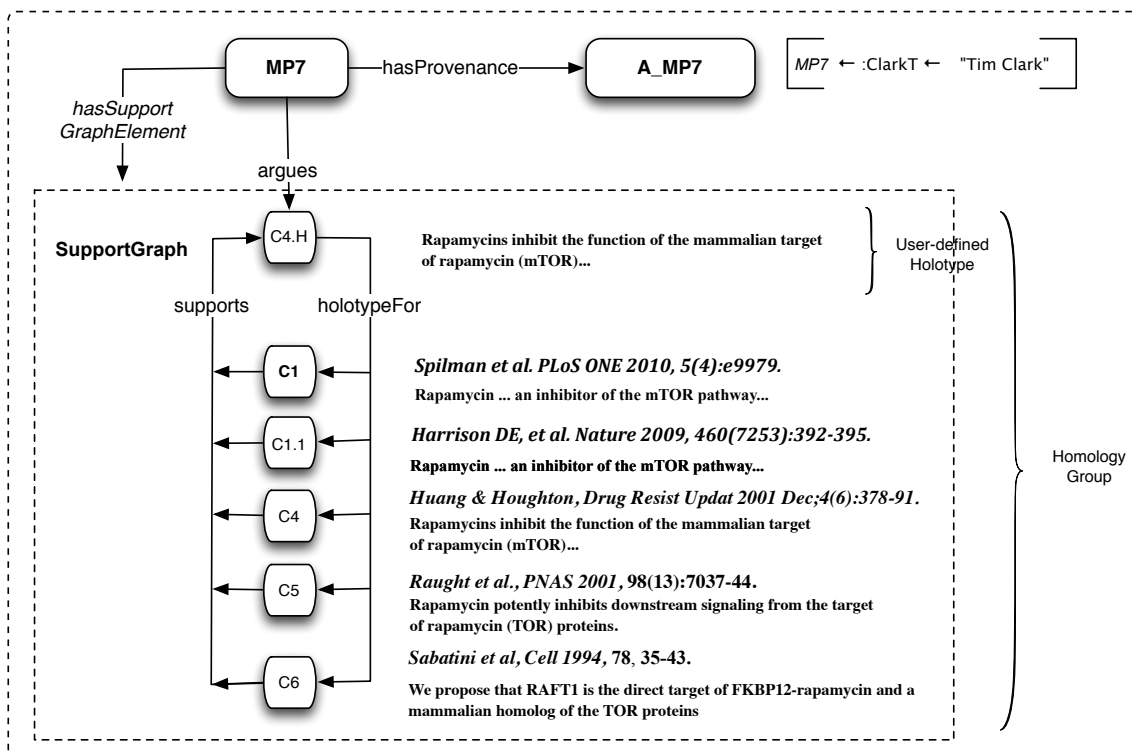

**Figure 13. A similarity group as a micropublication, with a user-defined holotype (representative) Claim. C4.H is the Claim. The other members of the similarity group are members of the holotype's SupportGraph, SG7. The Attribution of MP7, A_MP7, shows that the author and curator are the same person.**

In this case we model the BEL statement itself as a Micropublication. Here, the Argument Source for the BEL statement is a tuple from a relational database, consisting of the statement text and a reference to the source document, its PubMed ID.

Let $a_7$ =

    o    {"a(CHEBI:9168) =| kin(p(HGNC:FRAP1))" , "PMID: 12030785"}

We'll refer to the text of the BEL statement as C7, and to "PMID:12030785" as R96, for convenience.

Then for the corresponding micropublication MP7, we have

    o    the Claim is C7, "(CHEBI: 9168) =| kin(p(HGNC:FRAP1))", which in ordinary English means that sirolimus (also known as Rapamycin or CHEBI:9168) downregulates HUGO: FRAP1 (also known by its more common acronym, mTOR).

    o    A_C7, the Attribution of this Claim, is {"Pratt D, 2013"}, indicating that the Claim was formulated by Dexter Pratt;

    o    the SupportGraph is the support relation on {A_C7, R96}, indicating the reference to Huang and Houghton 200 describing the interaction of Rapamycin and mTOR in natural language.

As represented in **Figure 14**, the BEL statement references [89] as support. But this document-level reference can be resolved to a Claim-level reference as shown **Figure 15** and further to a network via its its similogs.

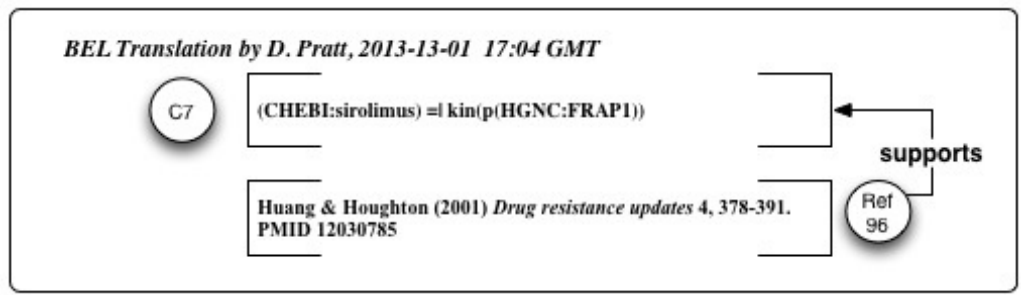

**Figure 14. BEL representation of a Claim from Spilman et al. 2010 [67] with support at document level only. Attribution of support is based on resolving the PubMed ID from a database.**

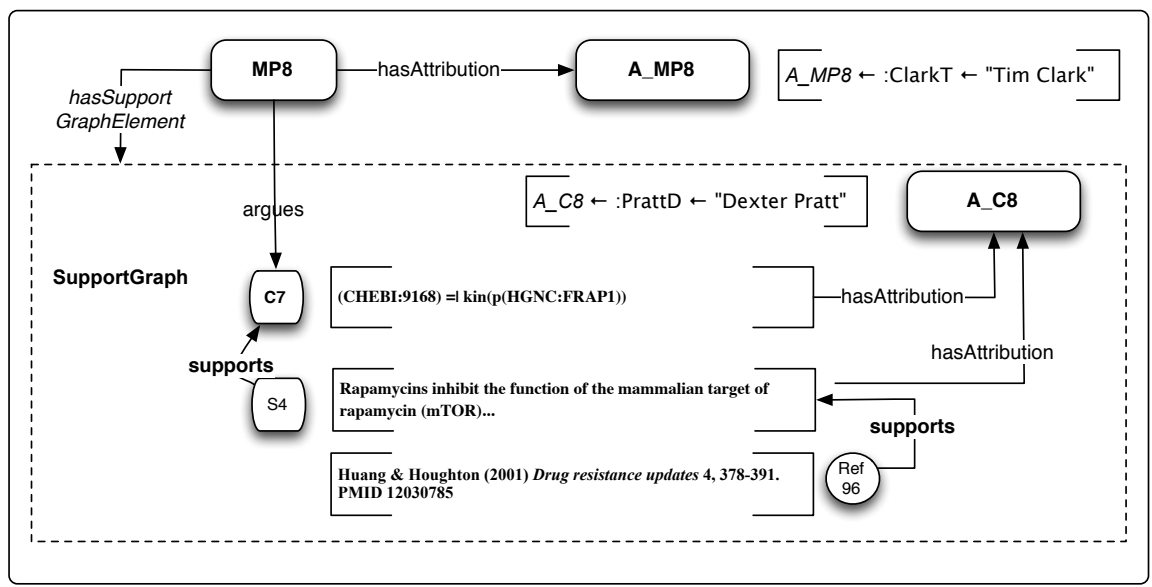

**Figure 15. BEL representation of a Claim from Spilman et al. 2010 [67] with support resolved to the Claim level. S4 is the specific Claim translated into the BEL statement.**

Claim-level resolution to similogs, would result in a BEL statement for the entire Similarity Group, asserting the Rapamycin ↔ mTOR inhibition interaction, whose support thus includes original evidence from the Sabatini group's publication as well as later review information. As the Claim must have been read and interpreted by a researcher to be translated at all, it seems clear that suitable software could embed specific citation to the Claim level at the time the asssociation with the BEL statement is databased.

BEL statements may also be modelled as nanopublications. An example of claim formalization in RDF using nanopublications is presented in **Section** Error! Reference source not found..

### 4.7 Example 7: Modeling Annotation and Discussion of Scientific Statements

_Use case_: Annotation is a key general use case that associates personal comments, discussion, semantic tags, or other constructs with scientific communication. Micropublications allow Annotations to be associated with logically explicit Backing, while contextualizing them (see

Section 0) within the digital content they describe. Readers as well as reviewers and other discussants may – with suitable software support – attach annotations directly to citable claims.

Let's suppose a reader or reviewer of the Spilman et al. article has some notes, comments, or observations to record – for example, about applicability of the Spilman result to drug discovery and development. The reader could then create annotation, modeled as a new, independent micropublication, referencing the original, as shown in **Figure 16**.

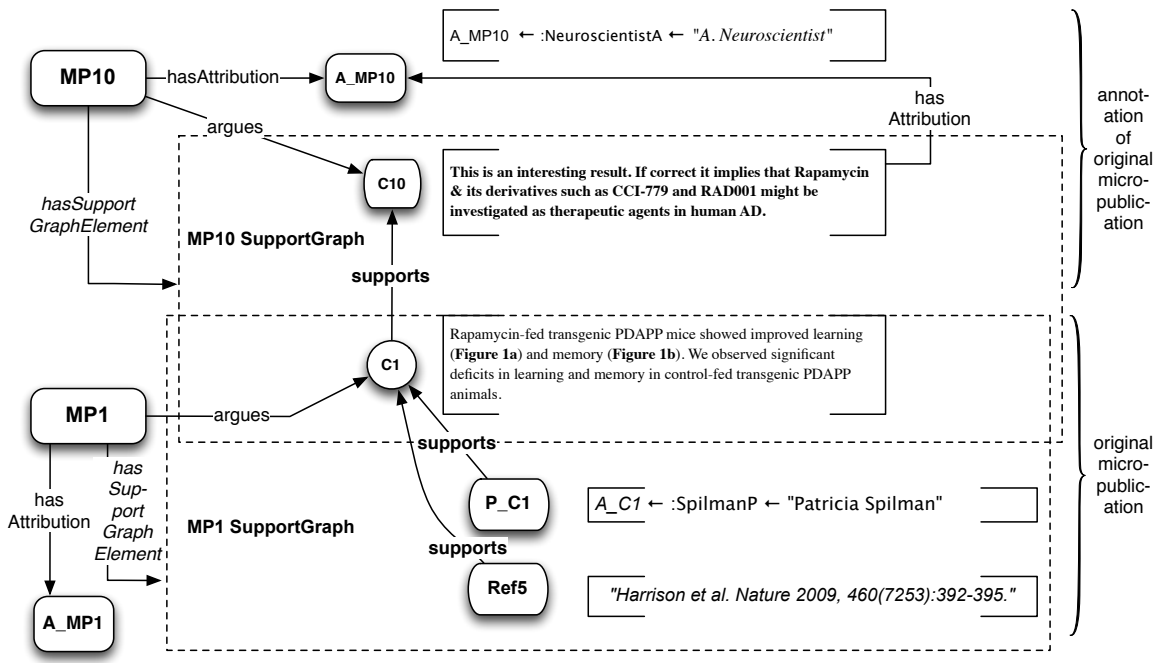

 **Figure 16. Annotation using Micropublication relations. C10 is a Claim made by the annotator, "A. Neuroscientist". It is supported by Statement C1 in Spilman et al. 2010, which it annotates.**

### 4.8    Example 8: Modeling Challenge and Disagreement

*Use case*: Scientific claims are defeasible. It is critical in evaluating Claims in biomedical research to review critical argumentation which may defeat them. Over time, not only Claims but Methods become established by withstanding such attacks.

As noted in Example 4, the the main Claim of [27] rests on three logical elements, the second of which is

- PDAPP mice are an established model of human Alzheimer Disease

by which the reader may reasonably infer, that this model is widely accepted as being adequate[d].

Spilman et al.'s principal Claim is still technically valid regardless of whether this is true or not, because the paper wisely restricts it to an assertion about the PDAPP mice only. But if in fact PDAPP mice do not model human Alzheimer's Disease well, the principal Claim as most readers would interpret it, is challenged. The whole point of the article is to suggest that Rapamycin or its analogs may be worthy of investigation as therapeutic agents for cases of actual human AD.

Bryan et al. [67], in their review of current transgenic mouse models of AD, make a number of critical points that potentially undercut the methodology used by [27]. One has to do with the low body temperatures of PDAPP transgenic mice. Low body temperatures can cause mice to become hypothermic in the Morris water maze (MWM) protocol; hypothermia impairs performance on the MWM, as previously diagrammed. Since this is an attack on the validity of the model of AD used in the Spilman et al. experiments, in constructing an extended network of Claims we would like to be alerted to this.

How do we model the challenge of Bryan et al. to the argument in Spilman et al.? In the SWAN project [20], we did so by asserting a formal relationship "inconsistentWith" between Claims. Doing so for any particular pair of Claims, was the task of the knowledge base curator. But the present model does not require a central curator - it is designed to support a collaborative ecosystem, which is a scalable model. So we adopt a different approach, in which annotations of inconsistency between micropublications, are themselves micropublications. This allows consistency/inconsistency relations to be asserted – and selectively consumed - at any point in the ecosystem.

There are two main ways the challenge relationship can be modeled, for two specific use cases.

The first way (Case 1) is to model the challenges made within a scientific article, against claims of another publication.

The second way (Case 2) is to model disagreement of two articles "from the outside", i.e. from the perspective of a reviewer or annotator. This approach can support as many different views as desired, of the relationships between different publications, and each view can be accorded its own attribution and authorship status, which can be selected for upon retrieval.

Note that *challenges* relationships are always part of the SupportGraph of the Micropublication in which they occur. Even challenges initiated elsewhere may be quoted and included, following Toulmin's approach to what he called *rebuttals*.

*Case 1: Micropublication models a Claim in one publication that challenges another.*

If an author of Publication A explicitly challenges a Claim of Publication B, that is a part of the discourse that can be modeled as part of the micropublication model of Publication A. In such a case, the "external" Claim being argued against will be quoted or summarized within Publication A.

Let $a_8$ = Bryan et al.'s review. Then for MP8 its micropublication, we have

> o  the Claim C11 from [67] as shown in **Figure 17**, "PDAPP mice tend to have lower body temperatures, which may result in varying degrees of hypothermia during the MWM task, which can produce amnesia in animals."
> o  A_C11 is the Attribution of C11;
> o  the SupportGraph is the set of support relations on {C11, R48, R49, R50 };
> o  the Claim C11 *challenges* statement S3 from MP3.

This would only be a good representation of Bryant et al., however, if the specific challenge to Spilman's paper was actually made in Bryant's text. In this case, it is not. So the observation must be made by a third party. We outline this situation in *Case 2*.

*Case 2: Micropublication models disagreement between two publications "from the outside".*

Suppose micropublications MP3 and MP11 are inconsistent, but it is a third party, not the author of either, who takes note of this fact. The initial micropublication summaries do not contain *challenge* relations, but the curator of a knowledge base (KB) wishes to note the discrepancy.

We can create a new micropublication, MP12, as follows.

Let a$_{12}$ =
- o     A (new) textual assertion that MP11:C11 and MP3:S3 are in conflict; with
- o     Summaries of the Claims MP11:C11 and MP3:S3.

Then for the MP12 formalization of a$_{12}$ we have

   o     the new assertion C12, an annotation by "KB Curator", stating: "Bryan et al. claim that PDAPP mice tend to have lower body temperatures, which may result in varying degrees of hypothermia during the MWM task, which in turn can produce amnesia in animals. This challenges the validity of PDAPP mice as an AD model, as asserted in Hsia et al."

This form is diagrammed in **Figure 18**.

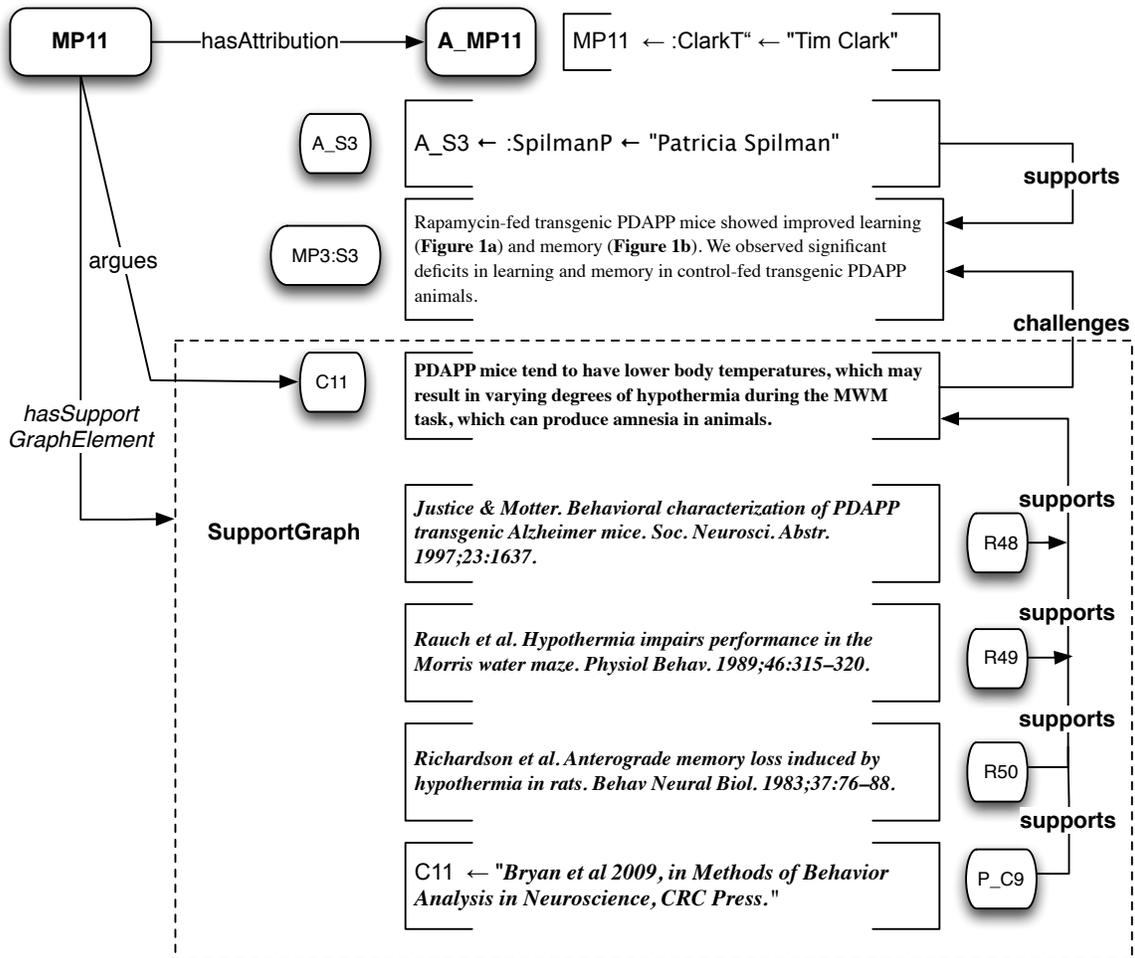

Figure 17. Claim C11, taken from [64], Bryan et al.'s review of transgenic mouse models of AD, challenges Statement MP3:S3 in Micropublication MP3. The Claim notes an issue with PDAPP mice tending to confound their performance in the Morris Water Maze (MWM). R48, R49 and R50 are document-level citations.

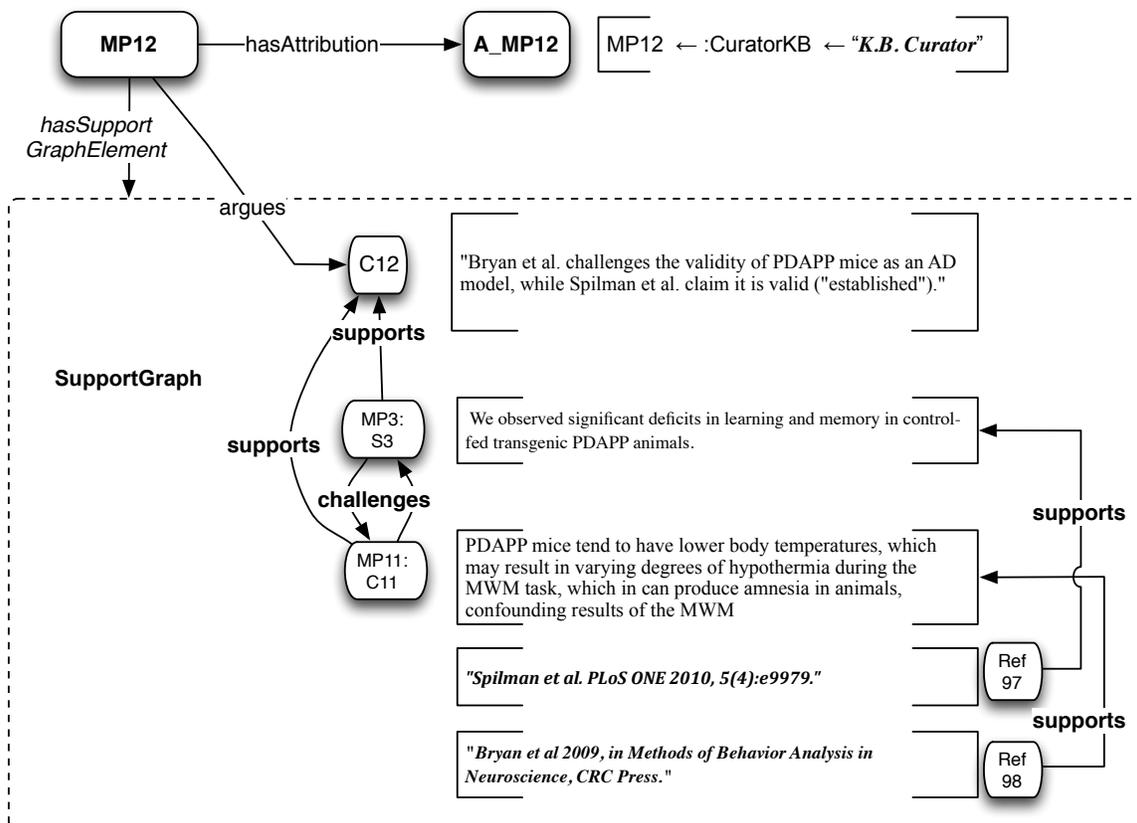

**Figure 18. Annotation of inconsistency between Spilman et al. 2010 and Bryan et al. 2009 as an independent micropublication. Challenge relationships between MP3:S3 and MP11:C11 are included in the SupportGraph SG12, because what MP12:C12 argues is precisely that these two Statements are in conflict.**

### 4.9  Example 9: Contextualization Using an Annotation Ontology

<u>Use cases</u>: (1) Making micropublications visually and experientially a part of the existing communications ecosystem for scientist users.  (2) Reliably mapping micropublication components to the documents from which they were extracted, or on which they were expressed as annotations, while (3) allowing them also to exist and be exchanged independently.

Biomedical researchers, and other workers in biomedical communications, spend a lot of time with the domain literature in their field.  They read it, write it, discuss it, annotate it, textmine it, give talks and journal clubs about it, and argue about it.  Our aim is to enhance and enable these practices, in a way that lets micropublications and other forms of annotation emerge as side-effects from improved practices. This requires (1) the ability to mash up annotations upon the literature they annotate, and (2) the ability to deconstruct annotations by direct reference to segments of existing literature.

We contextualize micropublications using an OWL ontology of annotations, which is orthogonal to the domain ontologies, and to the model of micropublications. Initially we used the Annotation Ontology, AO [56, 57]. We have now transitioned to a richer model, OA (the Open Annotation Model) [80], developed by the W3C Open Annotations Community Group, of which we were founding members[f].  However, the basic principles of these models are roughly the same.

AO and OA both allow free text, social tagging, or semantic tagging of Web documents using stand-off annotation. The AO and OA ontologies are vocabularies for describing what is annotated, what part of it is annotated (what segments or fragments), by whom and with what other provenance it was annotated, versioning, and what is the annotation associated with the target (text, ontology URIs, URI of another document, etc.).

"Stand-off annotation" means that the annotation is stored separately and references the documents via a special set of mechanisms. In the case of AO, these are called "selectors". A selector is a member of a class of mechanisms which specialise selection of document fragments by MIME type. For example, the text of ordinary HTML documents would use one type of selector, for example one specifying prefix, target, postfix and range. An image selector would work differently, using geometric specifications. In OA it is also possible to use W3C media fragments [93] as selectors. We have demonstrated, using AO, that annotation created on HTML target documents can be exported as RDF, and referenced independently by a PDF viewer (Utopia) [94, 95], which successfully mashed it up again into the correct text positions [96].

**Figure 19** shows a Claim in a Micropublication contextualized within a full-text article using vocabulary elements from the Open Annotations model.

In addition to the ontologies for annotation contextualization, which can be used by any application, we understand that useful document annotation tools are needed. Our team developed the Domeo web document annotation toolkit [17, 55, 58], now in version 2 release, for this purpose. Domeo is an open source licensed product (under Apache 2.0) and several groups are now collaborating with us on its development.

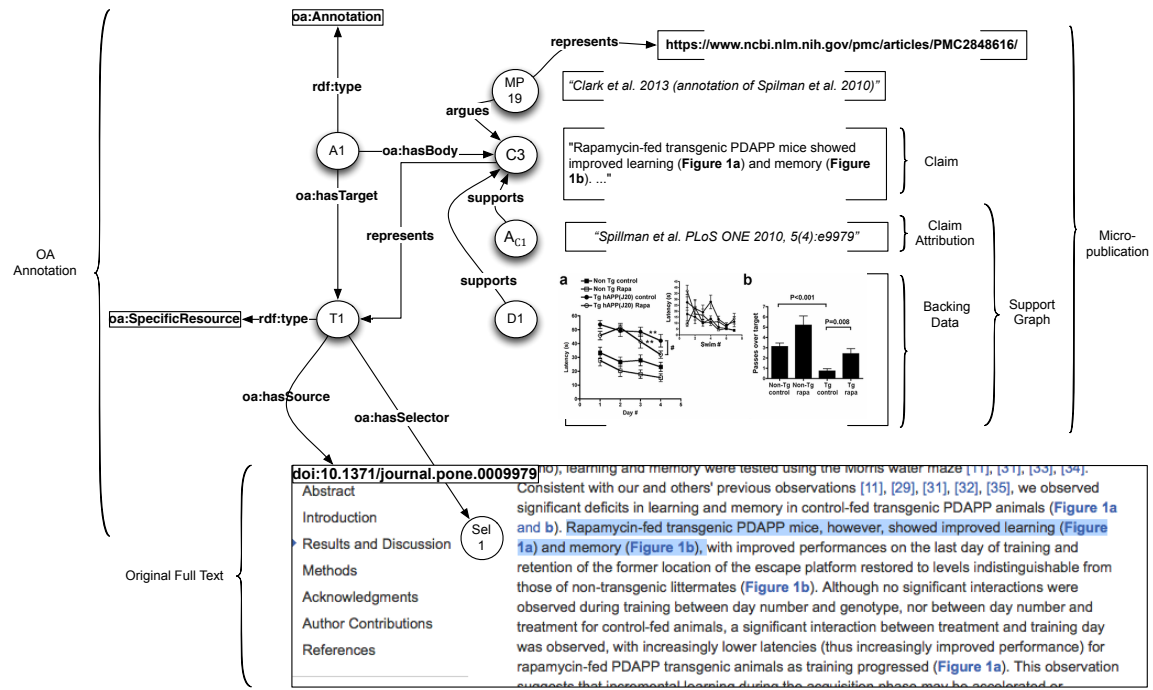

**Figure 19: Contextualizing a Claim within a full-text document using the Open Annotation ontology.**

## 5    Discussion

### 5.1    Supporting Reproducibility in Research

Greenberg's publications on citation distortion [3, 4] mentioned in the introduction, have great relevance to the problem of reproducibility of results. Begley and Ellis [8] mention two potential major sources of reproducibility problems,

- Cherry picking data, and
- Failure to properly describe methods.

They do not mention

- citation distortion, of various kinds

which Greenberg showed remains a common and unaddressed problem.

By providing a metadata model that can cite claims directly, micropublications enable citation networks to be constructed using a single common model, which will tend to distribute the costs. This should allow much more clarity to be brought to the question, "on what grounds is this statement made?" Furthermore, the networks may be resolved to purported foundational data and methods.

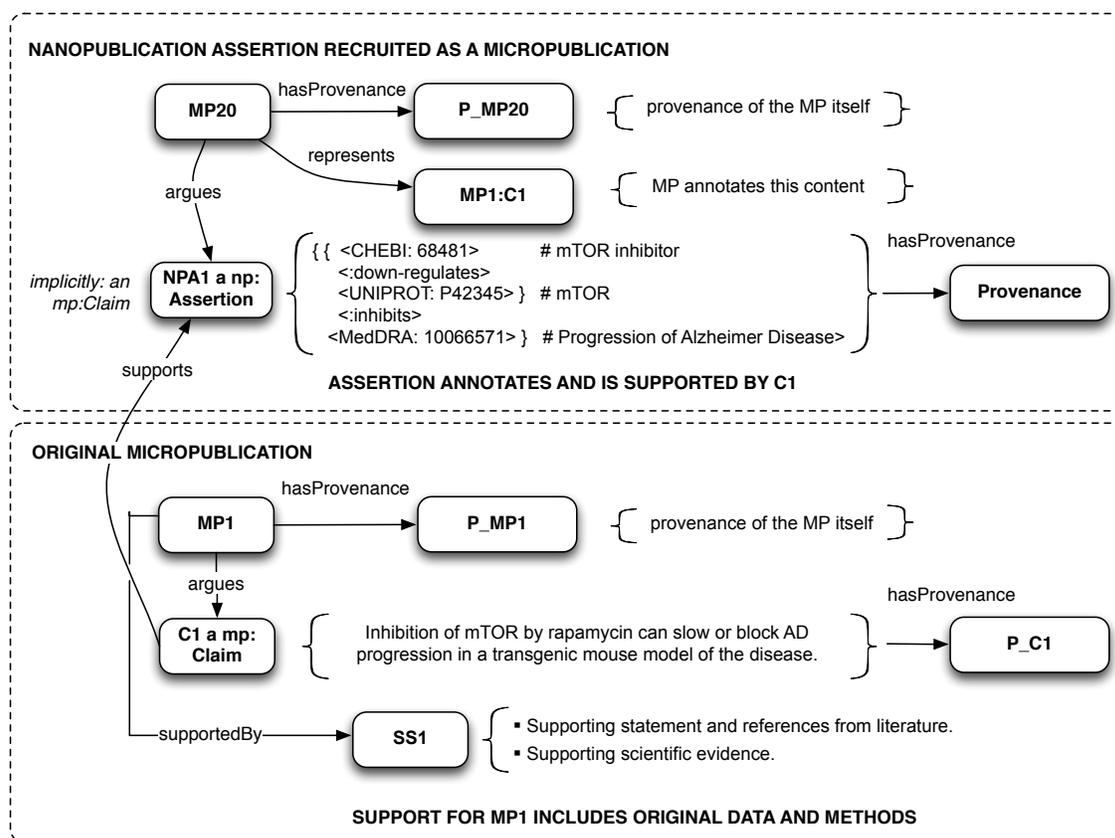

**Figure 20. A Nanopublication** `np:Assertion` **"recruited" as an** `mp:Claim` **and encapsulated within a Micropublication.**

**Figure 21** shows an example of citation distortion based on a section of Greenberg's supplemental data, represented as a micropublication citation networks drawn from eight publications [97-104].

It can be readily seen that Needham et al.'s [97] claim that "[t]he accumulation of APP and its fragments is often stated to precede other abnormalities in IBM muscle fibers…", made in a respected journal, is based only on citations to one laboratory, which repeatedly self cites, and whose supposed foundational references are nothing but hypotheses, if that. In fact, these authors (excluding the third) wrote another review published that same year in *Lancet Neurology*, which removed the "is often stated to" qualifier, upgrading the claim to "fact", without citation.

Greenberg also notes, in reviewing one of the "foundational" publications from the same laboratory[100], "major technical weaknesses" including "lack of quantitative data" and "lack of specificity of reagents". Both of these problems, particularly that of poor reagent specificity, can be highlighted readily in micropublications, because of the ability to cite both data and methods (reagents).

It is true that reagent citation is a more complex problem than simply providing and implementing this model can resolve. It needs urgent action by publishers to require more specificity in textual identification by authors, along with availability of global registries such as those provided by the Neuroscience Information Framework [58, 59, 105], for a reagent-citing approach to bear fruit. Our contribution here, is to provide an integrated approach to linking reagents and data directly and lucidly to the scientific claims they support.

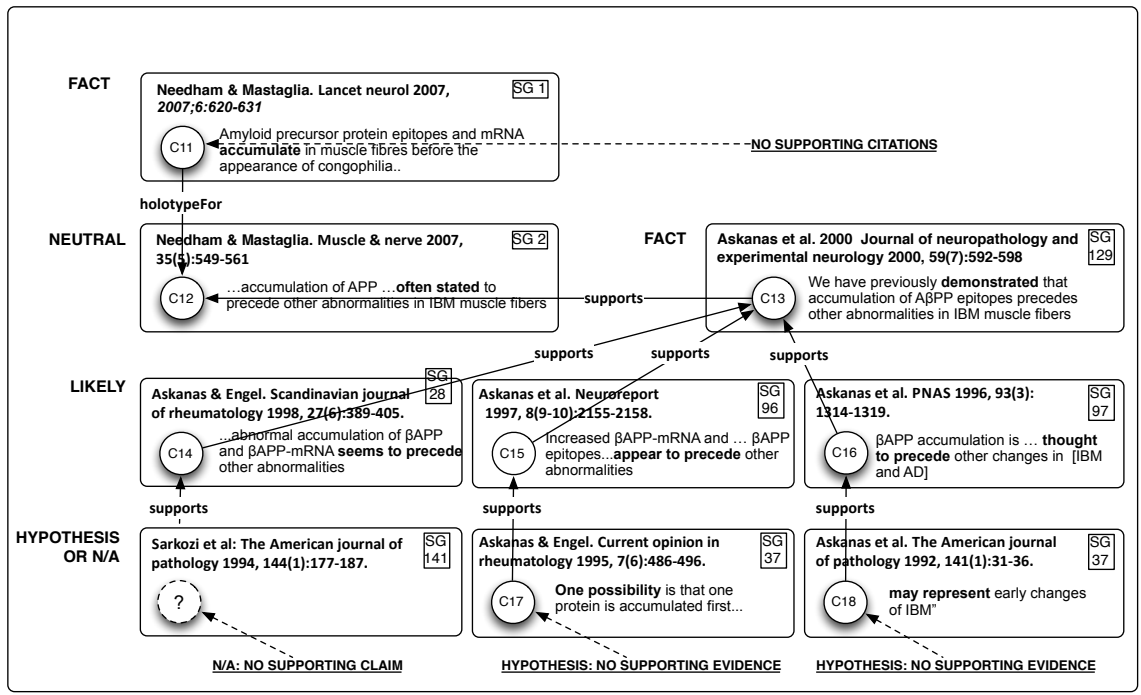

**Figure 21. Graph of a claim network from Greenberg 2009 [3]. The claim presented in this lineage states that amyloid beta deposition in IBM muscle fibers precedes other pathological changes – from which can be inferred that it is the causative factor. The foundational publications, which would be expected to contain supporting data, do not. Needham and Massaglia's article from Lancet Neurology provides no support at all for Claim C11 at all, treating it as a pure fact. In the Claim network, we assert C11 as a Holotype, or representative Statement, for C12. In fact, that assertion could be made for all Claims in the network.**

### 5.2    Implementation in Software

A natural question that arises is, can we embed this model usefully in software? This is required if it is to be effective. A related project in our group, Domeo [17, 55, 58], allows us to do so. Domeo is a web tookit for atuomated, semi-automated and manual annotation of web documents. It consists of a knowledgebase, parsers, web services, proxy server and browser-based interface. It has a plug-in based architecture and supports profile-based selection of user interfaces. Domeo is designed to support multiple knowededge-base instances communicating peer-to-peer, and allows annotations to be kept private, group-specific, or public.

Domeo is in active use as part of the Neuroscience Information Framework, and is installed for use by drug-hunting teams at a major pharmaceutical company. It is open source software licensed under Apache 2.0. It has a growing network of contributors, from both academia and industry.

Domeo version 1 allowed users to annotate web documents using the SWAN ontology of scientific discourse. As noted previously, one of the motivations for developing the Micropublications model was to address shortcomings in SWAN by means of a systematic re-thinking of the model, based on a different starting point.

We have now also implemented a Micropublications-based annotation functionality as an alpha plugin to Domeo version 2. The user interface (UI) for this plugin is shown in **Figure 22**, using the example Spilman et al. 2010 example again. A user begins by defining individual Statements and their support. This is done by highlighting the desired Statement, clicking "annotate" and selecting "micropublications" as the type of annotation. The Statement then appears in the panel and the user is given a choice of connecting it via the supportedBy relation to (a) its References, which have already been parsed out as computable objects; (b) data, in the form of images in the article, also already parsed out as computable and annotatable objects; or (c) other micropublication-annotated Statements within the article.

Currently Domeo implements an internal version of this model which it can serialize and store in a MongoDB noSQL database; this version is currently deployed for early alpha testing.

*One of the key features this model will provide in the context of pharmaceutical research, is the ability to link proprietary internal data as supporting evidence.*

We proposed this model as a general purpose representation model for claims, evidence and argument in the biomedical and other scientific literature. By design it is not intended to be limited in applicability to a single application. To be most useful it requires reasonably widespread adoption. In addition to the Domeo implementation we describe above, we suggest two particularly attractive points of uptake to the community at large.

1. Any bibliographic management software can implement micropublications. This would enable the scientist user to have a structured library at hand, not only of the references s/he has cited, but also of the particular statements within the referenced material, which are most important.
2. Publishers may implement micropublications as value-add annotation to their published content.

The main value proposition for implementing this model is to enable individual statements in the literature to be cited and referenced directly both by traditional articles and in web discussions,; linked easily into citation networks; and grounded in their supporting evidence – data and methods. We believe this is a pressing need across the biomedical research and development community, to provide better accountability and to support a culture of greater transparency. A number of unhelpful scholarly games might also be undermined by widespread adoption of our model, embedded in useful applications. Ultimately we believe it can greatly facilitate data and methods re-usability, improving the reliability and reproducibility of research results.

**Figure 22. Creating a micropublication in a Domeo version 2 alpha UI plugin. The micropublication *asserts* a Claim, which was simply selected by highlighting a few lines of full text from the original publication [27]. Support is provided by the cited Reference, which was retrieved and instantiated as an annotatable object when Domeo loaded the citing publication. Additional support is provided by the images of data shown, which was also loaded and instantiated on load of the containing publication. Original raw datasets, if cited and stored in a stable repository, may also provide support.**

### 5.3   Relationship to the SWAN Model

It may be helpful to outline some of the principal differences and correspondences between MP, and SWAN.

- Hypotheses and Micropublications
  - A `swan:Hypothesis` corresponds most closely to an `mp:Micropublication` together with the `Claim` it argues.
- Claims and Statements
  - A `swan:Claim` roughly corresponds to an `mp:Statement`.
- Evidence
  - SWAN models evidence for Claims in the form of literature citations, while Micropublications also model Data and Methods.

- o Micropublications support both literature citations-as-support (document level) as well as direct Claim / Statement citations (statement level).
- Statement Consistency
  - o In SWAN, a knowledgebase curator asserts inconsistency, between Claims globally, at a single point.
  - o In Micropublications, inconsistency is modeled as *challenge*, and may be asserted by anyone. There may or may not be a global curator, as needed.
- Statement Similarity
  - o SWAN implemented the notion of statement similarity using "canonical statements".
  - o Micropublications models statement similarity using similarity groups and uses the holotype of such a group as a surrogate claim to represent the common meaning.
  - o Micropublications similarity groups may be asserted as Micropublications themselves.
- Multi-polarity
  - o SWAN does not model multipolar argumentation. The concept of inconsistency is vague, relatively subjective, and global.
  - o Micropublications models multipolar argumentation, using challenge relationships. It is therefore compatible with bipolar argumentation frameworks. Furthermore, its *challenge* and *support*s relationships are explicitly published with attribution and authority, enabling multiple viewpoints to coexist and be selected/deselected.
- Abstraction and Annotation
  - o SWAN and Micropublications both support Comments.
  - o Micropublications supports translation of Claims into more or less any useful formal language capable of representing them in the required domain.
  - o Micropublications re-formalizations of Claims in other micropublications are also micropublications.
- Layering
  - o SWAN does not have formal mechanisms to support layering.
  - o Micropublications explicitly support layering. A Micropublications may be constructed about another Micropublications, and this may be layered as deeply as one wishes.
- Grounding
  - o The Micropublications model is theoretically grounded in Argumentation Theory. SWAN is not.

## 6   Conclusions

Micropublications enable us to formalize the arguments and evidence in scientific publications. We have shown the need for this kind of formalization in order to meet a number of significant use cases in the biomedical communications ecosystem. We have also demonstrated that purely statement-based models are too underfeatured to deal with scientific controversy and evidentiary requirements, and therefore cannot adequately meet use cases requiring an examination of evidence – which we believe will be the majority of those arising from the primary literature.

Nonetheless we believe statement-based models have a role in formalizing statements which have already been deployed with an adequate set of supporting evidence. These should also record the original authors' textual interpretation of the empirical evidence as the grounds for later formalization.

An example of this kind of application would be in drug hunting teams in a pharma or biotech company. Examination of evidence and possible contradiction is absolutely fundamental to the activities of such teams in qualifying targets and leads. At the same time, with adequate evidence chains, there is most certainly a role for statement formalization, and we have provided for this as an important interface in our model.

We have also attempted to show how the model not only meets various core use cases, but that it can be applied across a spectrum of complexity, from very simple annotations of the kind researchers make nearly every day using reference management and annotation software, to complex curated knowledgebases. This is a basic requirement for success in adoption: you don't have to buy the whole package. There is a simple entry point beyond which you need go no further. But you may go on, if you so choose.

We assume that implementation will be in various kinds of software environments in which the actual detailed construction of micropublications is built in to useful activities, where the formal instantiations of micropublications are constructed internally and behind the scenes – but may be shared at will. Our research group has built a micropublications capability into the DOMEO web annotation platform as a first step, and an evaluation of the model in this context will be presented in a forthcoming article.

As a general point, we expect value to accrue in the use of this model by enabling interoperability and incremental value addition through annotation, by users at various points in the scientific communication's ecosystem, who receive significant added utility in return for any additional work they put in to modeling activities. In use cases such as bibliographic management tools, much of the required modeling work is already done in everyday use of the existing tools. What is added (claim identification in cited works and specification of representative or holotype claims), is certainly going to be useful to the users who implement it.

The most promising approaches overall, will generate micropublications as standardized annotation metadata that can be exchanged and accumulated between applications.

As pointed out in [106], when annotation metadata is published using with the W3C Open Annotation Model [107] and includes semantic tags on the statements, it becomes a first-class object on the Web, and can be published, if desired, as linked data. This will not always be desirable due to privacy concerns for certain kinds of annotations and comments, and/or licensing issues. But in many cases we believe it would be useful.

We hope that other researchers will see the utility of one or more elements of this model for their own use cases of choice, and apply it. Developers and architects who wish to explore further uses of this model, including modifications and new applications, are invited to contact and discuss with the authors.

## Competing Interests

The authors declare no competing interests.

## Author Contributions

Tim Clark conducted the research, developed the micropublication use cases and designed the abstract model. He is the primary author of the OWL vocabulary and RDF examples based on the abstract model, and of this publication as a whole.

Paolo Ciccarese helped to check the consistency of the abstract model, and participated in developing the OWL vocabulary and the RDF examples. He also developed the Domeo Micropublications plugin.

Carole Goble supervised the research, and critiqued and edited several versions of this publication, including the present one.

### Acknowledgements

We are grateful for the support of Elsevier Laboratories, Eli Lilly and Company, the U.S. National Insitutes of Health (through the Neuroscience Information Framework) and anonymous donor foundations, which funded our work at the Massachusetts General Hospital.

Many thanks to our colleagues Bradley Allen, Anita Bandrowski, Judith Blake, Phil Bourne, Suzanne Brewerton, Monica Byrne, Christine Chichester, Anita De Waard, Michel Dumontier, Yolanda Gil, Paul Groth, Brad Hyman, Susan Kirst, Derek Marren, Maryann Martone, Barend Mons, Steve Pettifer, Dexter Pratt, Eric Prud'hommeaux, Marco Roos, Uli Sattler and Nigam Shah for numerous helpful discussions.

We also wish to thank the anonymous reviewers at J Biomed Semantics, for their close reading and careful critique of the manuscript. Our work benefitted substantially from their suggestions.

Part of this work was conducted using the Protégé resource, which is supported by grant GM10331601 from the National Institute of General Medical Sciences of the United States National Institutes of Health.

## Endnotes

[a] Today it would be more correct to say, "an information technology", but Shapin was writing before the advent of the Web.

[b] W.V.O. Quine criticized Strawson's formulation for similar reasons. [108]

[c] Thanks to our colleague Dexter Pratt for the BEL formulation of Rapamycin ↔ mTOR interaction.

[d] However, note that the author does not actually say "these mice are a good model of AD".

[e] Chapters VI-VIII of *On the Origin of Species* [109] provide a classic example of this form.

[f] All three of the current authors participated in the work of the W3C group, and the second author co-chaired it. Overall the W3C Open Annotation Community Group has had the participation of 63 institutions, with 113 participating individuals and representatives at the time of writing.